\documentclass[vecphys]{svmult}

\usepackage{makeidx}         
\usepackage{graphicx}        
\usepackage{multicol}        
\usepackage[bottom]{footmisc}

\makeindex             

\begin{document}

\title*{Lecture Notes on Semiconductor Spintronics}
\author{Tomasz Dietl}
\institute{Institute of Physics, Polish Academy of Sciences
and ERATO Semiconductor Spintronics Project, Japan Science and Technology Agency,
al. Lotnik\'ow 32/46, PL-02668 Warszawa, Poland
 and Institute of Theoretical Physics, Warsaw University, Poland
 \texttt{dietl@ifpan.edu.pl}}
%
%
\maketitle


{\bf Abstract:} These informal lecture notes describe the progress in semiconductor spintronics in a historic perspective as well as in a comparison to achievements of spintronics of ferromagnetic metals. After outlining motivations behind spintronic research, selected results of investigations on three groups of materials are presented. These include non-magnetic semiconductors, hybrid structures involving semiconductors and ferromagnetic metals, and diluted magnetic semiconductors either in paramagnetic or ferromagnetic phase. Particular attention is paid to the hole-controlled ferromagnetic systems whose thermodynamic, micromagnetic, transport, and optical properties are described in detail together with relevant theoretical models.

\section{Why Spintronics?}
\label{sec:1}
The well-known questions fuelling a broad interest in nanoscience are: will it still be possible to achieve further progress in information and communication technologies simply by continuing to miniaturize the transistors in microprocessors and the memory cells in magnetic and optical discs? How to reduce power consumption of components in order to save energy and to increase battery operation time? How to integrate nowadays devices with biological molecules and functionalities?

Since 70s, the miniaturization by obeying Moore's law has persistently lead to an exponential
increase in the quantity of information that can be processed, stored, and  transmitted  per unit area of microprocessor, memory, and fiberglass,  respectively. A modern integrated circuit contains now one billion transistors,  each smaller than 100~nm  in size, i.e., a five hundred times smaller than the  diameter of a human hair. The crossing of this symbolic 100~nm threshold at the  outset of the 21st century ushered in the era of nanotechnology. As the size of  transistors decreases, their speed increases, and their price falls. Today it is  much less expensive to manufacture one transistor than to print a single letter.
Despite the series of successes that industrial laboratories have scored over the  past 40 years in surmounting one technical and physical barrier after another,  there is a prevalent sense that in the near future a qualitative change is now in  store for us in terms of the methods of data processing, storing, encoding, and  transmission. For this reason, governments in many countries are financing  ambitious interdisciplinary programs aimed at insuring active participation in  the future development of nanotechnology.

Among the many proposals for where to take such research, the field of  spintronics, i.e., electronics aimed at understanding electron spin phenomena and  at proposing, designing, and developing devices to harness these phenomena, is  playing a major role. The hopes placed in spintronics are founded on the  well-known fact that since magnetic monopoles do not exist, random magnetic  fields are significantly weaker than random electric fields. For these reasons,  magnetic memories are non-volatile, while memories based on an accumulated  electric charge (dynamic random access memory, or DRAM) require frequent refreshing.

One of the ambitious goals in the spintronics field is to create magnetic random  access memory (MRAM), a type of device that would combine the advantages of both  magnetic memory and dynamic random access memory. This requires novel methods of  magnetizing memory cells and reading back the direction of such magnetization \index{magnetization},  which would not involve any mechanical systems. Another important step along this  path would be the ability to control magnetization \index{magnetization} isothermally, by means of  light or electric field. Modern devices expend relatively large amounts of energy  on controlling magnetization \index{magnetization} (\emph{i.e.}, storing data), as they employ Oersted magnetic  fields generated by electric currents.

The development of more "intelligent" magnetization \index{magnetization} control methods would also  make it possible to build spin transistors, devices composed of two layers of  ferromagnetic conductors separated by non-magnetic material. It stands to reason  that if carriers injected into the non-magnetic layer preserve their spin  direction, then the electric conductivity depends on the relative direction of  the magnetization \index{magnetization} vectors in the ferromagnetic layers. This could offer a means  of producing an energy-conserving and fast switching device, as it would allow  current to be controlled without changing the carrier concentration. An obvious  prerequisite for such a transistor to operate is the efficient injection of  spin-polarized carriers made of ferromagnetic material into the non-magnetic  area. Also, there should be no processes that could disrupt the spin  polarization. Simultaneously, researchers are seeking ways of generating, amplifying, and detecting spin currents: here, the underlying conviction is that the movement of electrons with opposite spins does not entail any losses, yet can carry information. This would lay the foundations for the development of low-power devices, characterized by significantly reduced heat dissipation. Another important issue is to develop  methods for injecting spin-polarized carriers into semiconductors. Apart from the  possibility of designing the magnetization \index{magnetization} sensors and spin  transistors, polarized carrier injection could prove to be useful as a method for  the fast modulation of semiconductor lasers and would allow surface-emission  lasers to work in a single mode fashion.

Perhaps the most important intellectual challenge to be faced in spintronics is  to create a hardware for quantum information science. Researchers over the world have joined  efforts to lay the theoretical foundations for this new discipline \cite{Benett:2000_a}, one notable  example being the Horodecki family from Gda\'nsk \cite{Horodecki:2003_a}. Experiments conducted by David  Awschalom's group in Santa Barbara show that spin degrees of freedom are of  particular importance as they maintain their phase coherence significantly longer  than orbital degrees of freedom do \cite{Kikkawa:1997_a}. Electron spin is therefore much more suitable  than electron charge for putting into practice modern ideas for performing  numerical computations using the superposition and entanglement of quantum states. Spin  nanostructures might consequently alter the basic principles not only in the  design of electronic elements, but also in the very computer architecture that  has been in use for half a century. It is noteworthy that quantum encoders are  already now being sold and installed: such devices use the polarization of light  to encode the transmitted information, and the unauthorized interception and  reading of this information appears to be impossible.

Today's research on spin electronics involves virtually all material families.  The most advanced are studies on magnetic multilayers. As demonstrated in 80s by groups of Albert Fert \cite{Baibich:1988_a} in Orsay and Peter Gr\"unberg \cite{Binasch:1989_a} in J\"ulich, these systems exhibit giant magnetoresistance (GMR  \index{giant magnetoresistance}). According to theory triggered by these discoveries and developed by J\'ozef Barna\'s from Pozna\'n and co-workers \cite{Camley:1989_a}, GMR  \index{giant magnetoresistance} results from spin-dependent scattering at adjacent interfaces between non-magnetic and magnetic metals, which changes when the magnetic field aligns magnetization \index{magnetization} of particular layers. Since 90s, the GMR  \index{giant magnetoresistance} devices have been successfully applied in reading heads of high-density hard-discs. Recent works focuss also on spin-dependent tunnelling via an oxide film. Remarkably, for the case of crystalline MgO sandwiched between contacts of amorphous Fe-Co-B layers, the difference between tunnelling resistance for anti-parallel and parallel orientations of magnetization \index{magnetization}, the TMR  \index{tunnelling magnetoresistance}, reaches a factor of three at 300~K \cite{Parkin:2004_a,Yuasa:2004_a,Ikeda:2005_a}. Moreover, the magnetization \index{magnetization} direction can be switched by an electric current below  $10^6$~A~cm$^{-2}$\cite{Hayakawa:2005_a}, opening the doors for a direct magnetization \index{magnetization} writing by current pulses.  Last but not least  such  structures can be used for injecting  highly polarized spin currents to semiconductors, such as GaAs \cite{Jiang:2005_a}.

These informal lecture notes on semiconductor spintronics exploit and update author's earlier reviews \cite{Dietl:1994_a,Dietl:2001_a,Dietl:2002_d,Dietl:2003_a,Dietl:2004_a,Dietl:2005_a,Dietl:2005_c}, where more systematic references to original papers can be found. Particular attention is paid here to those results of research on spin properties of semiconductors, which appear relevant in the context of disruptive classical and quantum information and communication technologies.  First part of the paper shows briefly how spin effects specific to non-magnetic semiconductors can be exploited in spintronic devices. This is followed by a presentation of chosen properties of hybrid semiconductor/ferromagnetic metal structures. The main body of the paper is devoted to diluted magnetic semiconductors (DMS \index{diluted magnetic semiconductors}), especially to materials exhibiting the ferromagnetic order, as they combine complementary resources of  semiconductor materials and ferromagnetic metals.  Here, the fundamental research  problem is to identify the extent to which the methods that have been so  successfully applied to controlling the density and degree of spin polarization  of carriers in semiconductor structures might be employed to control the  magnetization \index{magnetization} magnitude and direction.  Apart from the  possibility of designing the aforementioned magnetoresistive sensors and spin aligners, ferromagnetic semiconductors are the materials of choice for spin current amplification and detection. Furthermore, their outstanding magnetooptical properties can be exploited for fast light modulation as well as optical isolators, perhaps replacing hybrid structures consisting of paramagnetic DMS \index{diluted magnetic semiconductors}, such as (Cd,Mn)Te, and a permanent magnet.

In the course of the years semiconductor spintronics has evolved into a rather broad research field. These notes are by no means exhaustive and, moreover, they are biased by author's own expertise. Fortunately, however, in a number of excellent reviews the issues either omitted or only touched upon here has been thoroughly elaborated in terms of content and references to the original papers. For instance, the progress in fabrication and studies of spin quantum gates of double quantum dots has been described by van Viel et al. \cite{vanWiel:2003_a}. A comprehensive survey on spin-orbit \index{spin-orbit interaction} effects and the present status of spin semiconductor transistors has been completed by \v{Z}uti\'c, Fabian, and Das Sarma \cite{Zutic:2004_a}. Finally, Jungwirth \emph{et al.} \cite{Jungwirth:2006_a} have reviewed various aspects of theory of (Ga,Mn)As and related materials. Excellent reviews on the entire semiconductor spintronics are also available \cite{Wolf:2001_a,Ohno:2002_b}.

\section{Non-magnetic Semiconductors}
\label{sec:2}
\subsection{Overview}
The beginning of spintronic research on non-magnetic semiconductors can be traced back to the detection of nuclear spin polarization in Si illuminated by circularly polarized light reported in late 60s by Georges Lampel at Ecole Polytechnique \cite{Lampel:1968_a}. Already this pioneering experiment involved phenomena crucial for semiconductor spintronics: (i) the spin-orbit \index{spin-orbit interaction} interaction that allows for transfer of orbital (light) momentum to spin degrees of freedom and (ii) the hyperfine interaction between electronic and nuclear spins. Subsequent experimental and theoretical works on spin orientation in semiconductors, carried out in 70s mostly by researchers around Ionel Solomon in Ecole Polytechnique and late Boris P. Zakharchenya in Ioffe Institute, were summarized in a by now classic volume \cite{Meyer:1986_a}.

More recently, notably David Awschalom and his co-workers first at IBM and then at Santa Barbara, initiated the use of time resolved optical magneto-spectroscopies that have made it possible to both temporally and spatially explore the spin degrees of freedom in a wide variety of semiconductor materials and nanostructures \cite{Awschalom:2002_a}. The starting point of this experimentally demanding technique is the preparation of spins in a particular orientation by optically pumping into selected electronic states. The electron spin then precesses in an applied or molecular magnetic field produced by electronic or nuclear spins. The precessing magnetic moment creates a time dependent Faraday rotation of the femtosecond optical probe. The oscillation and decay measure the effective Land\'e g-factor, the local magnetic fields, and coherence time describing the temporal dynamics of the optically injected spins.

Present spintronic activities focuss on two interrelated topics. The first is to exploit Zeeman splitting and spin-orbit \index{spin-orbit interaction} interactions for spin manipulation. To this category belongs, in particular, research on spin filters and detectors, on the Datta-Das transistor \cite{Zutic:2004_a}, on optical generation of spin currents \cite{Ganichev:2003_a} and on the spin Hall effect \cite{Sih:2005_b}.  The other topic is the quest for solid-state spin quantum gates that would operate making use of spin-spin exchange  \index{exchange} \cite{Loss:1998_a} and/or hyperfine interactions \cite{Kane:1998_a}. An important aspect of the field is a dual role of the interactions in question in non-magnetic semiconductors: from one hand they allow for spin functionalities, on the other they account for spin decoherence and relaxation, usually detrimental for spin device performance. This, together with isotope characteristics, narrows rather severely a window of material parameters at which semiconductor spin devices might operate.

\subsection{Spin relaxation and dephasing}

\begin{figure} \centering
\includegraphics*[width=90mm]{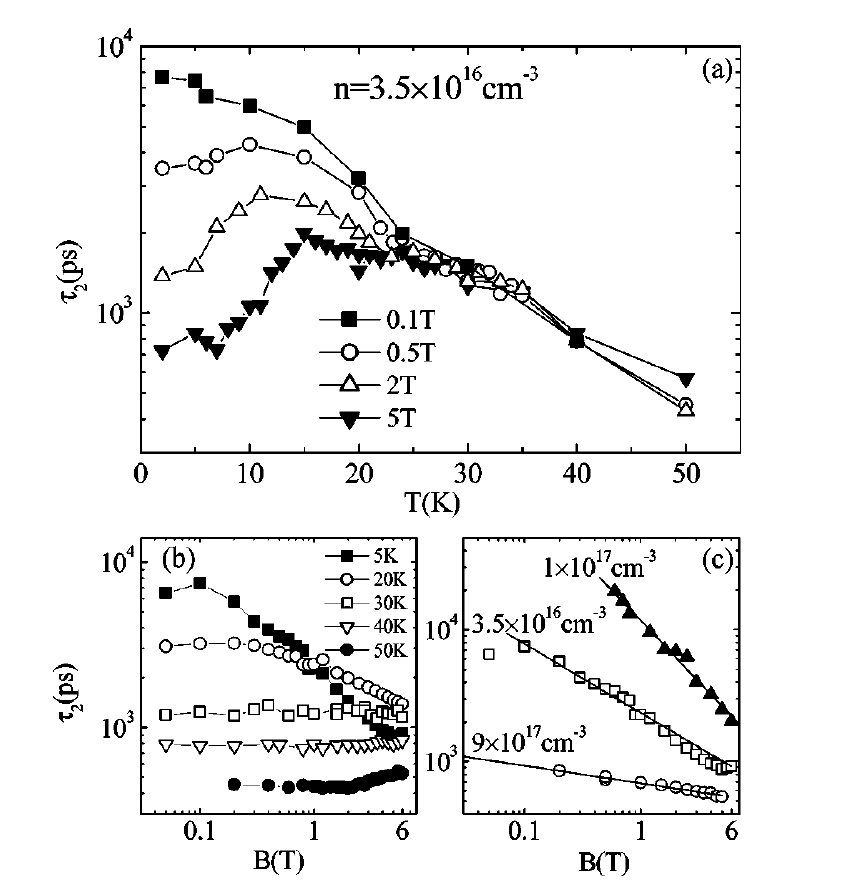}
\caption[]{Spin scattering time $\tau_2$ of n-GaN at various magnetic fields (a), temperatures (b)  ($n = 3.5\times 10^{16}$ cm$^{-3}$), and electron concentrations at 5 K (c) (after Beschoten {\it et al.} \cite{Beschoten:2001_a}).}
\label{fig:Beschoten}
\end{figure}

Owing to a large energy gap and the weakness of spin-orbit \index{spin-orbit interaction} interactions, especially long spin life times are to be expected in the nitrides and oxides. Figure 1 depicts results of time-resolved Faraday rotation, which has been used to measure electron spin coherence \index{spin coherence} in n-type GaN epilayers \cite{Beschoten:2001_a}. Despite densities of charged threading dislocations of $5\times 10^8$~cm$^{-2}$, this coherence yields spin lifetimes of about 20~ns at temperatures of 5~K, and persists up to room temperature.

\begin{figure} \centering
\includegraphics*[width=90mm]{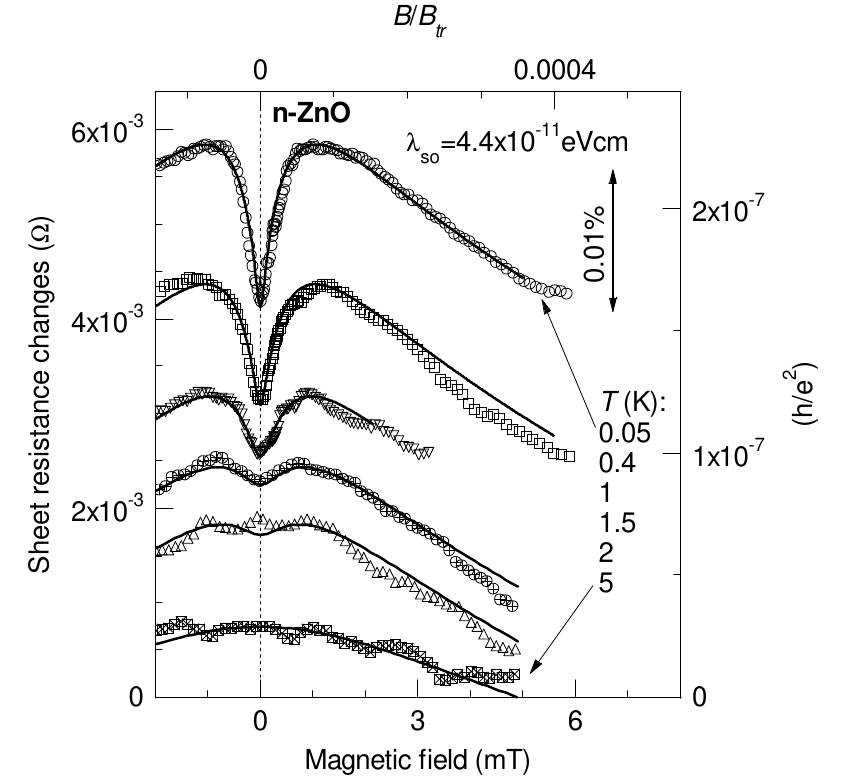}
\caption{Resistance changes in the magnetic field for
n-ZnO (symbols) compared to calculations (solid lines) within the
weak localization \index{weak localization} theory for the 2D case. Curves are vertically
shifted for clarity (after Andrearczyk {\it et al.} \cite{Andrearczyk:2005_c}).}
\label{fig:zno0wf}
\end{figure}

Figure~\ref{fig:zno0wf} presents a comparison of experimental and
calculated magnetoresistance (MR) of a ZnO:Al thin film containing $1.8\cdot 10^{20}$ electrons per cm$^{3}$ \cite{Andrearczyk:2005_c}. Here, spin effects control quantum interference corrections to the classical Drude-Boltzmann conductivity. A characteristic positive component of MR, signalizing the presence of spin-orbit \index{spin-orbit interaction} scattering, is detected below 1~mT at low temperatures.
This scattering is linked to the presence of a Rashba \index{Rashba hamiltonian}-like term
$\lambda_{\mbox{\scriptsize so}} \vec{c}(\vec{s}\times \vec{k})$ in
the $kp$ hamiltonian of the wurzite structure, first detected
in n-CdSe in the group of the present author \cite{Sawicki:1986_a}.
As shown in Fig.~\ref{fig:zno0wf}, a
quite good description of the findings is obtained with
$\lambda_{\mbox{\scriptsize so}} = 4.4\cdot 10^{-11}$~eV~cm,
resulting in the spin coherence \index{spin coherence} time  1~ns, more than $10^4$ times longer
than the momentum relaxation time. Importantly, this low decoherence rate of wide-band gap semiconductors is often coupled with a small value of the dielectric constant that enhances characteristic energy scales for quantum dot charging as well as for the exchange  \index{exchange} interaction of the electrons residing on the neighboring dots. This may suggest some advantages of these compounds for fabrication of spin quantum gates. Another material appealing in this
context is obviously Si, and related quantum structures, in which the interfacial electric field controls the magnitude of the Rashba \index{Rashba hamiltonian} term \cite{Wilamowski:2003_a} and material containing no nuclear spins can be obtained.

\subsection{An example of spin filter}

\begin{figure} \centering
\includegraphics*[width=90mm]{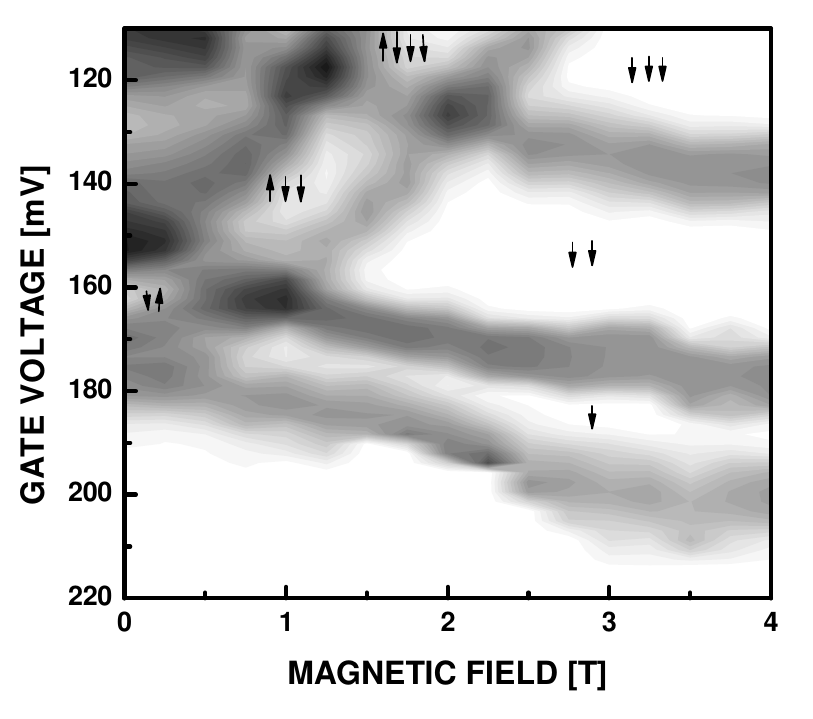}
\caption{Transconductance $dG/dV_g$ (gray scale) showing
dependence of 1D subbands on the magnetic field and gate voltage
for PbTe nanoconstriction of a wide
(Pb,Eu)Te/PbTe/(Pb,Eu)Te quantum well
(after Grabecki {\em et al.} \cite{Grabecki:2002_a}).}
\label{fig8}
\end{figure}

Turning to the case of narrow-gap semiconductors we note that
strong spin-orbit \index{spin-orbit interaction} effects specific to these
systems results, among other things, in a large Zeeman splitting of the carrier states,
which can be exploited for
fabrication of efficient spin filters. As an example, we consider
quantum point contacts patterned of PbTe
quantum wells embedded by Bi-doped Pb$_{0.92}$Eu$_{0.08}$Te
barriers \cite{Grabecki:2002_a,Grabecki:2005_a}.
Owing to biaxial strain, the fourfold L-valley degeneracy of the conduction
band in PbTe is lifted, so that the relevant ground-state 2D
subband is formed of a single valley with the long axis parallel
to the [111] growth direction. As discussed recently \cite{Grabecki:2005_a},
the paraelectric character of PbTe results in efficient screening
of Coulomb scattering potentials, so that signatures of ballistic
transport can be observed despite of significant amount of charged
defects in the vicinity of the channel. At the same time, the
electron density can be tuned over a wide range by biasing a p-n
junction that is formed between the p$^+$ interfacial layer and
the n-type quantum well \cite{Grabecki:2005_a}. Furthermore, a rather large magnitude of
electron spin splitting for the magnetic field along the growth
direction, corresponding to the Land\'e factor $|g*| \approx 66$,
can serve to produce a highly spin-selective barrier. According
to results displayed in Fig.~3, spin-degeneracy of the quantized conductance steps
starts to be removed well below 1~T, so that it has become possible to
generate entirely polarized spin current carried by a number of 1D
subbands \cite{Grabecki:2002_a}.

\section{Hybrid Structures}
\label{sec:3}
\subsection{Overview}
The hybrid nanostructures, in which
both electric and magnetic field are spatially modulated, are
usually fabricated by patterning of a ferromagnetic metal on the
top of a semiconductor  or by inserting ferromagnetic
nanoparticles or layers into a semiconductor matrix. In such
devices, the stray fields can control charge and spin dynamics in
the semiconductor. At the same time, spin-polarized electrons in
the metal can be injected into or across the semiconductor
\cite{Johnson:2002_a,Prinz:1998_a}. Furthermore, the ferromagnetic neighbors may affect
semiconductor electronic states by the ferromagnetic proximity
effect even under thermal equilibrium conditions. Particularly
perspective materials in the context of hybrid structures appear to
be those elemental or compound ferromagnets which can be grown in
the same reactor as the semiconductor counterpart.

\subsection{Spin injection}

It is now well established that efficient spin injection \index{spin injection} from
a ferromagnetic metal to a semiconductor is possible provided that
semiconductor Sharvin resistance is comparable or smaller
than the difference in interface resistances for two spin orientations.
Often, to enhance the latter,
a heavily doped or oxide layer is inserted between the metal and
as-grown semiconductor. In this way, spin current reaching polarization tens percents has been
injected form Fe into GaAs \cite{Hanbicki:2003_a,Jiang:2005_a}. At the same time, it is still hard
to achieve TMR  \index{tunnelling magnetoresistance} above 10\% in Fe/GaAs/Fe trilayer
structures without interfacial layer \cite{Zenger:2004_a}, which may suggest that the
relevant Schottky barriers
are only weakly spin selective.

The mastering of spin injection \index{spin injection} is a necessary
condition for the demonstration
of the Datta-Das transistor \cite{Datta:1990_a}, often regarded
as a flag spintronic device. In this spin FET, the orientation of the spins flowing between
ferromagnetic contacts, and thus the device resistance, is controlled by the Rashba \index{Rashba hamiltonian} field
generated in the semiconductor by an electrostatic gate. Recently, a current
modulation up to 30\% by the gate voltage was achieved in a Fe/(In,Ga)As/Fe FET
at room temperature \cite{Yoh:2005_a}. This important finding was obtained
for a 1~$\mu$m channel of narrow gap In$_{0.81}$Ga$_{0.19}$As, in which TMR  \index{tunnelling magnetoresistance} achieved 200\%, indicating
that the destructive role of the Schottky barriers
got reduced. Furthermore, an engineered interplay between the Rashba \index{Rashba hamiltonian} and Dresselhaus effects
\cite{Lusakowski:2003_a,Winkler:2004_a} resulted in a spin relaxation \index{spin relaxation} time long comparing to spin
precession period and the dwell time.

\subsection{Search for solid-state Stern-Gerlach effect \index{Stern-Gerlach effect}}

\begin{figure} \centering
\includegraphics*[width=120mm]{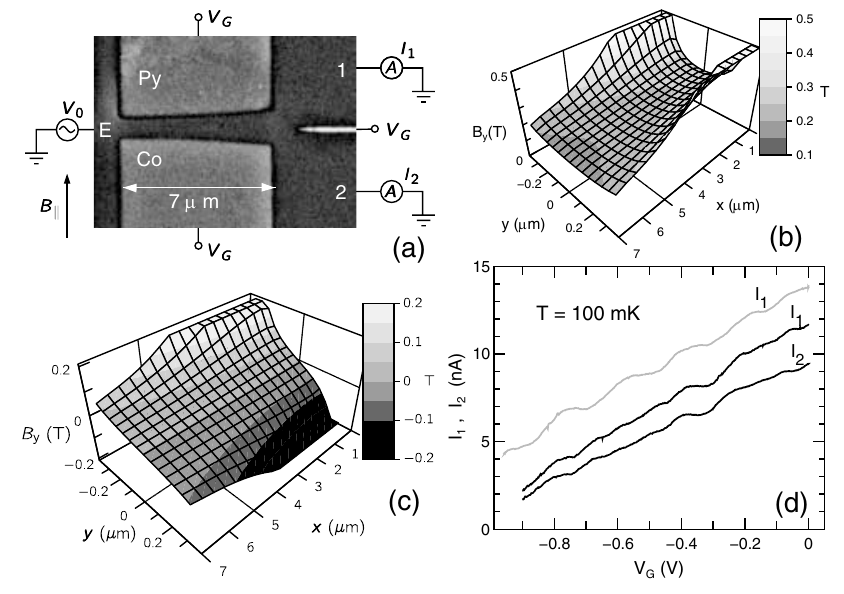}
\caption{(a) Scanning electron micrograph of the spin-filter
device. Fixed AC voltage $V_0$ is applied between emitter (E) and
"counters" (1), (2); $V_G$ is the DC gate voltage. The external
in-plane magnetizing field ($B_\parallel$) is oriented as shown.
(b) The in-plane magnetic field $B_y$ (wider part of the channel
is in front) calculated for half-plane, 0.1~$\mu$m thick magnetic
films separated by a position dependent gap $W(x)$ and magnetized
in the same directions (saturation magnetization \index{magnetization} as for Co).
(c)$B_y$ calculated for antiparallel directions of micromagnet
magnetization \index{magnetization}s. (d) Counter currents $I_1$ and $I_2$ as a function
of the gate voltage at $V_0=100$~$\mu$V and $B_\parallel = 0$;
upper curve (shown in gray) was collected during a different
thermal cycle and after longer infra-red illumination (after Wr\'obel
{\em et al.} \cite{Wrobel:2004_a}).} \label{Fig_1_SG}
\end{figure}

The ferromagnetic  component of hybrid structures can also serve for
the generation of a magnetic field. This field, if uniform,
produces a spin selective barrier that can serve as a local spin filter
and detector. A non-homogenous
field, in turn, might induce spatial spin separation via the Stern-Gerlach (S-G)
mechanism. Figure~\ref{Fig_1_SG}(a) presents a micrograph of a Stern-Gerlach device,
whose design results from an elaborated optimization
process \cite{Wrobel:2004_a}.  A local magnetic field was produced by NiFe
(permalloy, Py) and cobalt (Co) films. The micromagnets resided in deep groves
on the two sides of the wire, so that the 2D electron gas in the modulation-doped
GaAs/AlGaAs heterostructure was approximately at the center of the field, and
the influence of the competing Lorentz force was largely reduced.
Hall magnetometry was applied in order to visualize directly
the magnetizing process of the two micromagnets in question.

\begin{figure} \centering
\includegraphics*[width=90mm]{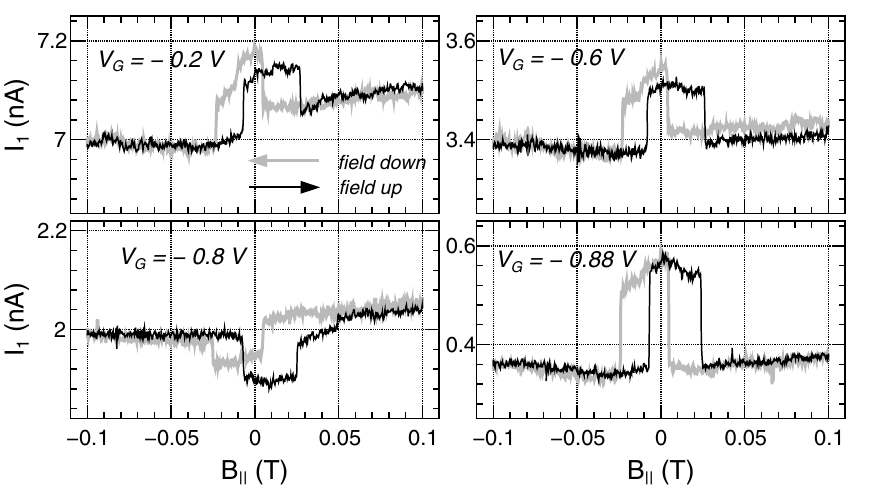}
\caption{The counter current $I_1$ of as a function of the in-plane
magnetic field for various gate voltages for the device shown
in Fig.~\ref{Fig_1_SG}. After~Wrobel {\em et al.} \cite{Wrobel:2004_a}.}
\label{Fig_2_SG}
\end{figure}

As shown in Fig.~\ref{Fig_2_SG}, a current increase in
counters was detected when a field gradient was produced by an appropriate cycle
of the external magnetic field at 100~mK. The range of magnetic fields where
the enhancement was observed corresponded to the the presence of the
field gradient according to the Hall magnetometry, which also
showed that Py magnetization \index{magnetization} diminished almost twofold prior to a
change in the direction of the external magnetic field. This
effect, associated with the formation of closure domains in soft
magnets, explained why the current changes appeared before the field
reversal. The relative change $\Delta I$ of counter current
depended on $V_G$, $\Delta I/I$ increased from $0.5$\% at zero gate
voltage to $50$\% close to the threshold. Furthermore, for $V_G$
about $-0.8$~V $\Delta I$ was negative. It was checked that
results presented in Fig.~\ref{Fig_2_SG} were unaltered by increasing
the temperature up to 200~mK and independent of the magnetic field
sweep rate.

Theoretical studies \cite{Wrobel:2004_a} of the results shown in Fig.~\ref{Fig_2_SG}
demonstrated that semiconductor nanostructures of the kind
shown in Fig.~\ref{Fig_1_SG} can indeed serve to generate and detect spin
polarized currents in the absence of an external magnetic field.
Moreover,  the degree and direction of
spin polarization at low electron densities can easily be
manipulated by gate voltage or a weak external magnetic field.
While the results of the performed computations suggest that the spin
separation and thus Stern-Gerlach effect \index{Stern-Gerlach effect} occurs under
experimental conditions in question, its direct experimental observation would
require incorporation of spatially resolved spin detection.

\section{Diluted Magnetic Semiconductors}
\label{sec:4}
\subsection{Overview}
This family of materials encompasses standard semiconductors, in which a sizable portion of atoms is substituted by such elements, which produce localized magnetic moments in the semiconductor matrix. Usually, magnetic moments originate from 3d or 4f open shells of transition metals or rare earths (lanthanides), respectively, so that typical examples of diluted magnetic semiconductors (DMS \index{diluted magnetic semiconductors}) are Cd$_{1-x}$Co$_x$Se, Ga$_{1-x}$Mn$_x$As \index{(Ga,Mn)As}, Pb$_{1-x}$Eu$_x$Te and, in a sense, Si:Er. A strong spin-dependent coupling between the band and localized states accounts for outstanding properties of DMS \index{diluted magnetic semiconductors}. This coupling gives rise to spin-disorder scattering, giant spin-splittings of the electronic states, formation of magnetic polarons, and strong indirect exchange  \index{exchange} interactions between the magnetic moments, the latter leading to collective spin-glass, antiferromagnetic or ferromagnetic spin ordering. Owing to the possibility of controlling and probing magnetic properties by the electronic subsystem or vice versa, DMS \index{diluted magnetic semiconductors} have successfully been employed to address a number of important questions concerning the nature of various spin effects in various environments and at various length and time scales. At the same time, DMS \index{diluted magnetic semiconductors} exhibit a strong sensitivity to the magnetic field and temperature as well as constitute important media for generation of spin currents and for manipulation of localized or itinerant spins by, e.g., strain, light, electrostatic or ferromagnetic gates. These properties, complementary to both non-magnetic semiconductors and magnetic metals, open doors for application of DMS \index{diluted magnetic semiconductors} as functional materials in spintronic devices.

Extensive studies of DMS \index{diluted magnetic semiconductors} started in 70s, particularly in the group of Robert R. Ga{\l}\c{a}zka in Warsaw, when appropriately purified Mn was employed to grow bulk II-VI Mn-based alloys by various modifications of the Bridgman method \cite{Galazka:1978_a}. Comparing to magnetic semiconductors, such as Eu chalcogenides (e.g., EuS) and Cr spinels (e.g., CdCr$_2$Se$_4$) investigated earlier \cite{Nagaev:1983_a}, DMS \index{diluted magnetic semiconductors} exhibited smaller defect concentrations and were easier to dope by shallow impurities. Accordingly, it was possible to examine their properties by powerful magnetooptical and magnetotransport techniques \cite{Dietl:1994_a,Galazka:1978_a,Dietl:1981_a,Furdyna:1988_a}. Since, in contrast to magnetic semiconductors, neither narrow magnetic bands nor long-range magnetic ordering affected low-energy excitations, DMS \index{diluted magnetic semiconductors} were named semimagnetic semiconductors. More recently, research on DMS \index{diluted magnetic semiconductors} have been extended toward materials containing magnetic elements other than Mn as well as to III-VI, IV-VI \cite{Bauer:1992_a}  and III-V \cite{Matsukura:2002_c} compounds as well as group IV elemental semiconductors and various oxides \cite{Prellier:2003_a}. In consequence, a variety of novel phenomena has been discovered, including effects associated with narrow-bands and magnetic phase transformations, making the borderline between properties of DMS \index{diluted magnetic semiconductors} and magnetic semiconductors more and more elusive.

A rapid progress of DMS \index{diluted magnetic semiconductors} research in 90s stemmed, to a large extend, from the development of methods of crystal growth far from thermal equilibrium, primarily by molecular beam epitaxy (MBE), but also by laser ablation. These methods have made it possible to obtain DMS \index{diluted magnetic semiconductors} with the content of the magnetic constituent beyond thermal equilibrium solubility limits \cite{Ohno:1998_a}. Similarly, the doping during MBE process allows one to increase substantially the electrical activity of shallow impurities \cite{Haury:1997_a,Ferrand:2001_a}. In the case of III-V DMS \index{diluted magnetic semiconductors} \cite{Matsukura:2002_c}, in which divalent magnetic atoms supply both spins and holes, the use of the low-temperature MBE (LT MBE) provides thin films of, e.g., Ga$_{1-x}$Mn$_x$As \index{(Ga,Mn)As} with $x$ up to 0.07 and the hole concentration in excess of $10^{20}$~cm$^{-3}$, in which ferromagnetic ordering is observed above 170~K \cite{Wang:2005_j}. Remarkably, MBE and processes of nanostructure fabrication, make it possible to add magnetism to the physics of semiconductor quantum structures. Particularly important are DMS \index{diluted magnetic semiconductors}, in which ferromagnetic ordering was discovered, as discussed in some details later on.

\subsection{Magnetic impurities in semiconductors}

A good starting point for the description of DMS \index{diluted magnetic semiconductors} is the Vonsovskii model, according to which the electron states can be divided into two categories: (i) localized magnetic d or f shells and (ii) extended band states built up of s, p, and sometimes d atomic orbitals. The former give rise to the presence of local magnetic moments and intra-center optical transitions. The latter form bands, much alike as in the case of non-magnetic semiconductor alloys. Indeed, the lattice constant of DMS \index{diluted magnetic semiconductors} obeys the Vegard low, and the energy gap $E_g$ between the valence and the conduction band depends on $x$ in a manner qualitatively similar to non-magnetic counterparts. According to the Anderson model, the character of magnetic impurities in solids results from a competition between (i) hybridization of local and extended states, which tends to delocalized magnetic electrons and (ii) the on-site Coulomb interactions among the localized electrons, which stabilizes the magnetic moment in agreement with Hund's rule.

\begin{figure} \centering
\includegraphics[width=130mm]{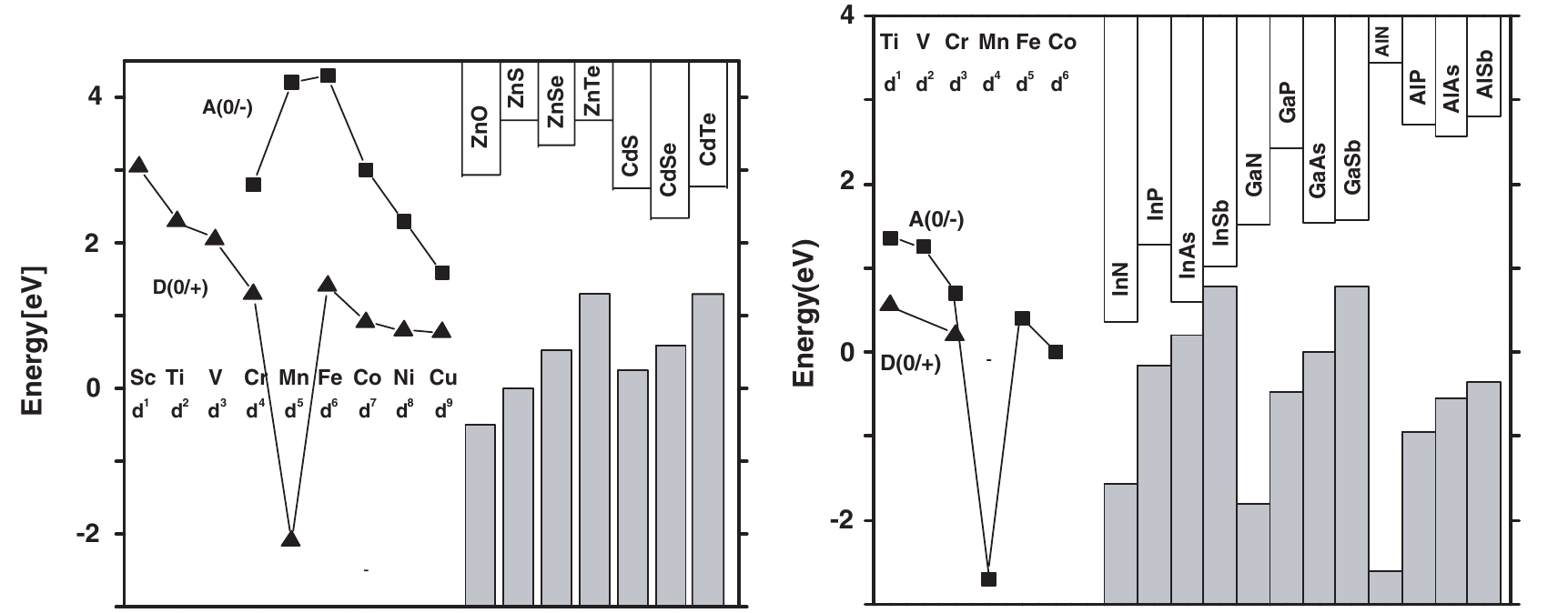}
\caption{Approximate  positions of transition metals levels
relative to the conduction and valence band edges of II-VI (left
panel) and III-V (right panel) compounds. By triangles the
d$^{N}$/d$^{N-1}$ donor and by squares the d$^N$/d$^{N+1}$ acceptor
states are denoted (adapted from Langer et al. \cite{Langer:1988_a} and Zunger \cite{Zunger:1986_a}).}
\label{Fig:1_SST}
\end{figure}

Figure \ref{Fig:1_SST} shows positions of local states derived from 3d shells of transition metal (TM) impurities in respect to the band energies of the host II-VI and III-V compounds. In Fig.The levels labelled "donors" denote the ionization energy of the magnetic electrons (TM$^{2+} \rightarrow$~TM$^{3+}$ or d$^{n} \rightarrow$~d$^{n-1}$), whereas the "acceptors" correspond to their affinity energy (TM$^{2+} \rightarrow$~TM$^{1+}$ or d$^{n} \rightarrow$~d$^{n+1}$). The difference between the two is the on-d-shell Coulomb (Hubbard) repulsion energy $U$ in the semiconductor matrix. In addition, the potential introduced by either neutral or charged TM can bind a band carrier in a Zhang-Rice-type singlet or hydrogenic-like state, respectively.  Such bound states are often experimentally important, particularly in III-V compounds, as they correspond to lower energies than the competing d-like states, such as presented in Fig.~\ref{Fig:1_SST}.

In the case of Mn, in which the d shell is half-filled, the d-like donor state lies deep in the valence band, whereas the acceptor level resides high in the conduction band, so that $U \approx 7$~eV according to photoemission and inverse photoemission studies. Thus, Mn-based DMS \index{diluted magnetic semiconductors} can be classified as charge transfer insulators, $E_g < U$.  The Mn ion remains in the 2+ charge state, which means that it does not supply any carriers in II-VI materials.  However, it acts as a hydrogenic-like acceptor in the case of III-V antimonides and arsenides, while the corresponding Mn-related state is deep, presumably due to a stronger p-d hybridization, in the case of phosphides and nitrides. According to Hund's rule the total spin $S = 5/2$ and the total orbital momentum $L = 0$ for the d$^5$ shell in the ground state. The lowest excited state d$^{*5}$ corresponds to $S = 3/2$ and its optical excitation energy is about 2~eV. Thus, if there is no interaction between the spins, their magnetization \index{magnetization} is described by the paramagnetic Brillouin function. In the case of other transition metals, the impurity-induced levels may appear in the gap, and then compensate shallow impurities, or even act as resonant dopant, e.g., Sc in CdSe, Fe in HgSe or Cu in HgTe. Transport studies of such systems have demonstrated that inter-site Coulomb interactions between charged ions lead to the Efros-Shklovskii gap in the density of the impurity states, which makes resonant scattering to be inefficient in semiconductors \cite{Wilamowski:1990_a}. Furthermore, spin-orbit \index{spin-orbit interaction} interaction and Jahn-Teller effect control positions and splittings of the levels in the case of ions with $L \ne 0$. If the resulting ground state is a magnetically inactive singlet there is no permanent magnetic moment associated with the ion, the case of Fe$^{2+}$, whose magnetization \index{magnetization} is of the Van Vleck-type at low temperatures.

\subsection{Exchange interaction between band and localized spins}
The important aspect of DMS \index{diluted magnetic semiconductors} is a strong spin-dependent coupling of the effective mass carriers to the localized d electrons, first discovered in (Cd,Mn)Te \cite{Komarov:1977_a,Gaj:1978_b} and (Hg,Mn)Te \cite{Bastard:1978_a,Jaczynski:1978_a}. Neglecting non-scalar corrections that can appear for ions with $L \ne 0$, this interaction assumes the Kondo form,

\begin{equation}
H_K = - I(\vec{r}-\vec{R}^{(i)})\vec{s}\vec{S}^{(i)},
\end{equation}
where $I(\vec{r}-\vec{R}^{(i)})$ is a short-range exchange  \index{exchange} energy operator between the carrier spin $\vec{s}$ and the TM spin localized at $\vec{R}^{(i)}$. When incorporated to the $kp$ scheme, the effect of $H_K$ is described by matrix elements  $\langle u_i|I|u_i\rangle$, where $u_i$ are the Kohn-Luttinger amplitudes of the corresponding band extreme. In the case of carriers at the $\Gamma$ point of the Brillouin zone in zinc-blende DMS \index{diluted magnetic semiconductors}, the two relevant matrix elements $\alpha =\langle u_c|I|u_c\rangle$  and  $\beta =\langle u_v|I|u_v\rangle$ involve s-type and p-types wave functions, respectively. There are two mechanisms contributing to the Kondo coupling \cite{Dietl:1981_a,Bhattacharjee:1983_a,Kacman:2001_a}: (i) the exchange  \index{exchange} part of the Coulomb interaction between the effective mass and localized electrons; (ii) the spin-dependent hybridization between the band and local states. Since there is no hybridization between $\Gamma_6$ and d-derived (e$_g$ and t$_{2g}$) states in zinc-blende structure, the s-d coupling is determined by the direct exchange  \index{exchange}. The experimentally determined values are of the order of $\alpha N_o \approx 0.25$~eV, where $N_o$ is the cation concentration, somewhat reduced comparing to the value deduced from the energy difference between $S±1$ states of the free singly ionized Mn atom 3d$^5$4s$^1$, $\alpha N_o  = 0.39$~eV. In contrast, there is a strong hybridization between $\Gamma_8$ and t$_{2g}$ states, which affects their relative position, and leads to a large magnitude of  $|\beta N_o| \approx 1$~eV. If the relevant effective mass state is above the  t$_{2g}$ level (the case of, e.g., Mn-based DMS \index{diluted magnetic semiconductors}), $\beta < 0$ but otherwise $\beta$ can be positive (the case of, e.g., Zn$_{1-x}$Cr$_x$Se \cite{Mac:1993_a}).

\subsection{Electronic properties}
\subsubsection{Effects of giant spin splitting}
In the virtual-crystal and molecular-field approximations, the effect of the Kondo coupling is described by
$H_K = I\vec{M}(\vec{r})\vec{s}/g\mu_B$, where $\vec{M}(\vec{r})$ is magnetization \index{magnetization} (averaged over a microscopic region around $\vec{r}$) of the localized spins, and $g$ is their Land\'e factor. Neglecting thermodynamic fluctuations of magnetization \index{magnetization} (the mean-field approximation) $\vec{M}(\vec{r})$ can be replaced by $M_o(T,H)$, the temperature and magnetic field dependent macroscopic magnetization \index{magnetization} of the localized spins available experimentally. The resulting spin-splitting of s-type electron states is given by
\begin{equation}
 \hbar \omega_s = g^*\mu_BB + \alpha M_o(T,H)/g\mu_B,
\end{equation}
where $g^*$ is the band Land\'e factor. The exchange  \index{exchange} contribution is known as the giant Zeeman splitting, as in moderately high magnetic fields and low temperatures it  attains values comparable to the Fermi energy or to the binding energy of excitons and shallow impurities. For effective mass states, whose periodic part of the Bloch function contains spin components mixed up by a spin-orbit \index{spin-orbit interaction} interaction, the exchange  \index{exchange} splitting  does not depend only on the product of $M_o$ and the relevant exchange  \index{exchange} integral, say $\beta$, but usually also on the magnitude and direction of  $\vec{M}_o$, confinement, and strain. Furthermore, because of confinement or non-zero $k$ the Bloch wave function contains contributions from both conduction and valence band, which affects the magnitude and even the sign of the spin splitting \cite{Bastard:1978_a,Jaczynski:1978_a,Furdyna:1988_a,Merkulov:1999_a}. The giant Zeeman splitting is clearly visible in magnetooptical phenomena as well as in the Shubnikov-de Haas effect, making an accurate determination of the exchange  \index{exchange} integrals possible, particularly in wide-gap materials, in which competing Landau and ordinary spin splittings are small.

The possibility of tailoring the magnitude of spin splitting in DMS \index{diluted magnetic semiconductors} structures offers a powerful tool to examine various phenomena. For instance, spin engineering was explored to control by the magnetic field the confinement of carriers and photons \cite{Sadowski:1997_a}, to map atom distributions at interfaces \cite{Gaj:1994_a} as well as to identify the nature of optical transitions and excitonic states. Furthermore, a subtle influence of spin splitting on quantum scattering amplitude of interacting electrons with opposite spins was put into evidence in DMS \index{diluted magnetic semiconductors} in the weakly localized regime in 3D \cite{Sawicki:1986_a}, 2D \cite{Smorchkova:1997_a,Jaroszynski:2005_b}, and 1D systems \cite{Jaroszynski:1995_a}. The redistribution of carriers between spin levels induced by spin splitting was found to drive an insulator-to-metal transition \cite{Wojtowicz:1986_a} as well as to generate universal conductance fluctuations in DMS \index{diluted magnetic semiconductors} quantum wires \cite{Jaroszynski:1995_a}. Since the spin splitting is greater than the cyclotron energy, there are no overlapping Landau levels in modulation-doped heterostructures of DMS \index{diluted magnetic semiconductors} in the quantum Hall regime in moderately strong magnetic fields. This made it possible to test a scaling behavior of wave functions at the center of Landau levels \cite{Jaroszynski:2000_a}. At higher fields, a crossing of Landau levels occurs, so that quantum Hall ferromagnet could be evidenced and studied \cite{Jaroszynski:2002_a}. At the same time, it has been confirmed that in the presence of a strong spin-orbit \index{spin-orbit interaction} coupling (e.g., in the case of p-type wave functions) the spin polarization can generate a large extraordinary (anomalous) Hall voltage \cite{Brandt:1984_a}. Last but not least, optically \cite{Oestreich:1999_a} and electrically controlled spin-injection \cite{Fiederling:1999_a} and filtering \cite{Slobodskyy:2003_a} were observed in all-semiconductor structures containing DMS \index{diluted magnetic semiconductors}.

\subsubsection{Spin-disorder scattering}
Spatial fluctuations of magnetization \index{magnetization}, disregarded in the mean-field approximation, lead to spin disorder scattering. According to the fluctuation-dissipation theorem, the corresponding scattering rate in the paramagnetic phase is proportional to $T\chi(T)$, where $\chi(T)$ is the magnetic susceptibility of the localized spins \cite{Dietl:1994_a,Dietl:1991_a}. Except to the vicinity of ferromagnetic phase transitions, a direct contribution of spin-disorder scattering to momentum relaxation turns out to be small. In contrast, this scattering mechanism controls the spin lifetime of effective mass carriers in DMS \index{diluted magnetic semiconductors}, as evidenced by studies of universal conductance fluctuations \cite{Jaroszynski:1998_a}, line-width of spin-flip Raman scattering \cite{Dietl:1991_a}, and optical pumping efficiency \cite{Krenn:1989_a}. Furthermore, thermodynamic fluctuations contribute to the temperature dependence of the band gap and band off-set. In the case when the total potential introduced by a magnetic ion is grater than the width of the carrier band, the virtual crystal and molecular field approximations break down, a case of the holes in Cd$_{1-x}$Mn$_x$S. A non-perturbative scheme was developed \cite{Benoit:1993_a,Dietl:1998_a} to describe nonlinear dependencies of the band gap on $x$ and of the spin splitting on magnetization \index{magnetization} observed in such situations.

\subsection{Magnetic polarons}
Bound magnetic polaron (BMP), that is a bubble of spins ordered ferromagnetically by the exchange  \index{exchange} interaction with an effective mass carrier in a localized state, modifies optical, transport, and thermodynamic properties of DMS \index{diluted magnetic semiconductors}. BMP is formed inside the localization radius of an occupied impurity or quantum dot state but also around a trapped exciton, as the polaron formation time is typically shorter than the exciton lifetime \cite{Dietl:1995_a}. The BMP binding energy and spontaneous carrier spin-splitting are proportional to the magnitude of local magnetization \index{magnetization}, which is built up by two effects: the molecular field of the localized carrier and thermodynamic fluctuations of magnetization \index{magnetization} \cite{Dietl:1982_a,Dietl:1983_a,Dietl:1983_b,Dietl:1994_a}. The fluctuating magnetization \index{magnetization} leads to dephasing and enlarges width of optical lines. Typically, in 2D and 3D systems, the spins alone cannot localized itinerant carriers but in the 1D case the polaron is stable even without any pre-localizing potential \cite{Benoit:1993_a}. In contrast, a free magnetic polaron---a delocalized carrier accompanied by a travelling cloud of polarized spins---is expected to exist only in magnetically ordered phases. This is because coherent tunnelling of quasi-particles dressed by spin polarization is hampered, in disordered magnetic systems, by a smallness of quantum overlap between magnetization \index{magnetization}s in neighboring space regions. Interestingly, theory of BMP can readily be applied for examining effects of the hyperfine coupling between nuclear spins and carriers in localized states.

\subsection{Exchange interactions between localized spins}
As in most magnetic materials, classical dipole-dipole interactions between magnetic moments are weaker than exchange  \index{exchange} couplings in DMS \index{diluted magnetic semiconductors}. Direct d-d or f-f exchange  \index{exchange} interactions, known from properties of magnetic dimmers, are thought to be less important than indirect exchange  \index{exchange} channels. The latter involve a transfer of magnetic information via spin polarization of bands, which is produced by the exchange  \index{exchange} interaction or spin-dependent hybridization of magnetic impurity and band states. If magnetic orbitals are involved in the polarization process, the mechanism is known as superexchange  \index{exchange}, which is merely antiferromagnetic and dominates, except for p-type DMS \index{diluted magnetic semiconductors}. If fully occupied band states are polarized by the sp-d exchange  \index{exchange} interaction, the resulting indirect d-d coupling is known as the Bloembergen-Rowland mechanism. In the case of Rudermann-Kittel-Kasuya-Yosida (RKKY) interaction, the d-d coupling proceeds via spin polarization of partly filled bands, that is by free carriers. Since in DMS \index{diluted magnetic semiconductors} the sp-d is usually smaller than the width of the relevant band (weak coupling limit) as well as the carrier concentration is usually smaller than those of localized spins, the energetics of the latter can be treated in the continuous medium approximation, an approach referred here to as the Zener model. Within this model the RKKY interaction is ferromagnetic, and particularly strong in p-type materials, because of a large magnitudes of the hole mass and exchange  \index{exchange} integral $\beta$. It worth emphasizing that the Zener model is valid for any ratio of the sp-d exchange  \index{exchange} energy to the Fermi energy. Finally, in the case of systems in which magnetic ions in different charge states coexist, hopping of an electron between magnetic orbitals of neighboring ions in differing charge states tends to order them ferromagnetically. This mechanism, doubted the double exchange  \index{exchange}, operates in manganites but its relevance in DMS \index{diluted magnetic semiconductors} has not yet been found.

    In general, the bilinear part of the interaction Hamiltonian for a pair of spins $i$ and $j$ is described by a tensor $\hat{J}$,

    \begin{equation}
     H_{ij} = -2\vec{S}^{(i)}\hat{J}^{(ij)}\vec{S}^{(j)},
    \end{equation}
which in the case of the coupling between nearest neighbor cation sites in the unperturbed zinc-blende lattice contains four independent components. Thus, in addition to the scalar Heisenberg-type coupling, $H_{ij} = -2J^{(ij)}\vec{S}^{(i)}\vec{S}^{(j)}$,  there are non-scalar terms (e.g., Dzialoshinskii-Moriya or pseudo-dipole). These terms are induced by the spin-orbit \index{spin-orbit interaction} interaction within the magnetic ions or within non-magnetic atoms mediating the spin-spin exchange  \index{exchange}. The non-scalar terms, while smaller than the scalar ones, control spin-coherence time and magnetic anisotropy \index{magnetic anisotropy}. Typically, $J^{(ij)} \approx -1$~meV for nearest-neighbor pairs coupled by the superexchange  \index{exchange}, and the interaction strength decays fast with the pair distance. Thus, with lowering temperature more and more distant pairs become magnetically neutral, $S_{tot} = 0$. Accordingly, the temperature dependence of magnetic susceptibility assumes a modified Curie form, $\chi(T) = C/T^{\gamma}$, where $\gamma  < 1$ and both $C$ and $\gamma$ depend on the content of the magnetic constituent $x$. Similarly, the field dependence of magnetization \index{magnetization} is conveniently parameterized by a modified Brillouin function $B_S$ \cite{Gaj:1979_a},

   \begin{equation}
    M_o(T,H) = Sg\mu_BN_ox_{eff}{\mbox{B}}_S[Sg\mu_BH/k_B(T + T_{AF})],
    \label{Eq:BF}
    \end{equation}
in which two $x$-- and $T$--dependent empirical parameters, $x_{eff} < x$ and $T_{AF} > 0$, describe the presence of antiferromagnetic interactions.

\subsection{Magnetic collective phenomena}

In addition to magnetic and neutron techniques \cite{Galazka:1995_a}, a variety of optical and transport methods, including $1/f$ noise study of nanostructures \cite{Jaroszynski:1998_a}, have successfully been employed to characterize collective spin phenomena in DMS \index{diluted magnetic semiconductors}. Undoped DMS \index{diluted magnetic semiconductors} belong to a rare class of systems, in which spin-glass freezing is driven by purely antiferromagnetic interactions, an effect of spin frustration inherent to the randomly occupied fcc sublattice.  Typically, in II-VI DMS \index{diluted magnetic semiconductors}, the spin-glass freezing temperature $T_g$ increases from 0.1~K for $x = 0.05$ to 20~K at $x = 0.5$ according to $T_g \sim x^{\delta}$, where $\delta \approx 2$, which reflects a short-range character of the superexchange  \index{exchange}. For $x$ approaching 1, antiferromagnetic type III ordering develops, according to neutron studies. Here, strain imposed by the substrate material--the strain engineering--can serve to select domain orientations as well as to produce spiral structures with a tailored period \cite{Giebultowicz:1992_a}. Particularly important is, however, the carrier-density controlled ferromagnetism of bulk and modulation-doped p-type DMS \index{diluted magnetic semiconductors} described next.

\section{Properties of Ferromagnetic Semiconductors}
\label{sec:6}
\subsection{Overview}
Since for decades III-V semiconductor compounds have been applied as photonic and microwave devices, the discovery of ferromagnetism first in In$_{1-x}$Mn$_x$As \cite{Ohno:1992_a} and then in Ga$_{1-x}$Mn$_x$As \index{(Ga,Mn)As} by Hideo Ohno and collaborators in Sendai \cite{Ohno:1996_a} came as a landmark achievement. In these materials, substitutional divalent Mn ions provide localized spins and function as acceptor centers that provide holes which mediate the ferromagnetic coupling between the parent Mn spins \cite{Dietl:1997_a,Matsukura:1998_a,Jungwirth:1999_a}.  In another technologically important group of semiconductors, in II-VI compounds, the densities of spins and carriers can be controlled independently, similarly to the case of IV-VI materials, in which hole-mediated ferromagnetism was discovered by Tomasz Story {\em et al.} in Warsaw already in the 80s \cite{Story:1986_a}. Stimulated by the theoretical predictions of the present author \cite{Dietl:1997_a}, laboratories in Grenoble and Warsaw, led by late Yves Merle d'Aubign\'e and the present author, joined efforts to undertake comprehensive research dealing with carrier-induced ferromagnetism in II-IV materials containing Mn. Experimental studies conducted with the use of magnetooptical and magnetic methods led to the discovery of ferromagnetism in 2D and \cite{Haury:1997_a} 3D II-VI materials \cite{Ferrand:2001_a} doped by nitrogen acceptors.

Guided by the growing amount of experimental results, the present author and co-workers proposed a theoretical model of the hole-controlled ferromagnetism in III-V, II-VI, and group IV semiconductors containing Mn \cite{Dietl:2000_a,Dietl:2001_b}.  In these materials conceptual difficulties of charge transfer insulators and strongly correlated disordered metals are combined with intricate properties of heavily doped semiconductors, such as Anderson-Mott localization and defect generation by self-compensation mechanisms. Nevertheless,  the theory built on Zener's model of ferromagnetism and the Kohn-Luttinger $kp$ theory of the valence band in tetrahedrally coordinated semiconductors has quantitatively described thermodynamic, micromagnetic, transport, and optical properties of DMS \index{diluted magnetic semiconductors} with delocalized or weakly localized holes \cite{Dietl:2000_a,Dietl:2001_b,Jungwirth:2006_a,Kepa:2003_a}, challenging competing theories. It is often argued that owing to these studies Ga$_{1-x}$Mn$_x$As \index{(Ga,Mn)As} has become one of the best-understood ferromagnets. Accordingly, this material  is now employed as a testing ground for various  \emph{ab initio} computation approaches to strongly correlated and disordered systems. Moreover, the understanding of the carrier-controlled ferromagnetic DMS \index{diluted magnetic semiconductors} has provided a basis for the development of novel methods enabling magnetization \index{magnetization} manipulation and switching.

\subsection{p-d Zener model}

   It is convenient to apply the Zener model of carrier-controlled ferromagnetism by introducing the functional of free energy density, ${\cal{F}}[\vec{M}(\vec{r})]$. The choice of the local magnetization \index{magnetization} $\vec{M}(\vec{r})$ as an order parameter means that the spins are treated as classical vectors, and that spatial disorder inherent to magnetic alloys is neglected. In the case of magnetic semiconductors ${\cal{F}}[\vec{M}(\vec{r})]$ consists of two terms, ${\cal{F}}[\vec{M}(\vec{r})] = {\cal{F}}_S[\vec{M}(\vec{r})] + {\cal{F}}_c[\vec{M}(\vec{r})]$, which describe, for a given magnetization \index{magnetization} profile $\vec{M}(\vec{r})$, the free energy densities of the Mn spins in the absence of any carriers and of the carriers in the presence of the Mn spins, respectively. A visible asymmetry in the treatment of the carries and of the spins corresponds to an adiabatic approximation: the dynamics of the spins in the absence of the carriers is assumed to be much slower than that of the carriers. Furthermore, in the spirit of the virtual-crystal and molecular-field approximations, the classical continuous field $\vec{M}(\vec{r})$ controls the effect of the spins upon the carriers. Now, the thermodynamics of the system is described by the partition function $Z$, which can be obtained by a functional integration of the Boltzmann factor $\exp(-\int d\vec{r}{\cal{F}}[\vec{M}(\vec{r})]/k_BT)$ over all magnetization \index{magnetization} profiles $\vec{M}(\vec{r})$ \cite{Dietl:1983_a,Dietl:1983_b}. In the mean-field approximation (MFA), a term corresponding to the minimum of ${\cal{F}}[\vec{M}(\vec{r})]$ is assumed to determine $Z$ with a sufficient accuracy.

If energetics is dominated by spatially uniform magnetization \index{magnetization} $\vec{M}$, the spin part of the free energy density in the magnetic field $\vec{H}$ can be written in the form \cite{Swierkowski:1988_a}
\begin{equation}
{\cal{F}}_S[\vec{M}] = \int_0^{\vec{M}} d \vec{M}_o\vec{h}(\vec{M}_o) - \vec{M}\vec{H}.
\end{equation}
Here, $\vec{h}(\vec{M}_o)$ denotes the inverse function to $\vec{M}_o(\vec{h})$, where $\vec{M}_o$ is the available experimentally macroscopic magnetization \index{magnetization} of the spins in the absence of carriers in the field $h$ and temperature $T$. In DMS \index{diluted magnetic semiconductors}, it is usually possible to parameterize $M_o(h)$ by the Brillouin function that, according to Eq.~\ref{Eq:BF}, takes the presence of intrinsic short-range antiferromagnetic interactions into account. Near $T_C$ and for $H = 0$, $M$ is sufficiently small to take $M_o(T,h) = \chi(T)h$, where $\chi(T)$ is the magnetic susceptibility of localized spins in the absence of carriers. Under these conditions,
\begin{equation}
{\cal{F}}_S[M] = M^2/2\chi(T),
\end{equation}
which shows that the increase of ${\cal{F}}_S$ with $M$ slows down with lowering temperature, where $\chi(T)$ grows. Turning to ${\cal{F}}_c[M]$ we note that owing to the giant Zeeman splitting of the bands proportional to $M$, the energy of the carriers, and thus ${\cal{F}}_c[M]$, decreases with $|M|$, ${\cal{F}}_c[M] -{\cal{F}}_c[0]\sim -M^2$. Accordingly, a minimum of ${\cal{F}}[M]$ at non-zero $M$ may develop in $H = 0$ at sufficiently low temperatures signalizing the appearance of a ferromagnetic order.

The present authors and co-workers \cite{Dietl:2000_a} found that the minimal hamiltonian necessary to describe properly effects of the complex structure of the valence band in tetrahedrally coordinated semiconductors upon ${\cal{F}}_c[M]$ is the Luttinger  $6\times 6$ $kp$  model supplemented by the p-d exchange  \index{exchange} contribution taken in the virtual crystal and molecular field approximations,
\begin{equation}
H_{pd} = \beta \vec{s}\vec{M}/g\mu_B.
\end{equation}
This term leads to spin splittings of the valence subbands, whose magnitudes---owing to the spin-orbit \index{spin-orbit interaction} coupling---depend on the hole wave vectors $\vec{k}$ in a complex way even for spatially uniform magnetization \index{magnetization} $\vec{M}$. It would be technically difficult to incorporate such effects to the RKKY model, as the spin-orbit \index{spin-orbit interaction} coupling leads to non-scalar terms in the spin-spin Hamiltonian. At the same time, the indirect exchange  \index{exchange} associated with the virtual spin excitations between the valence subbands, the Bloembergen-Rowland mechanism, is automatically included. The model allows for biaxial strain, confinement, and was developed for both zinc blende and wurzite materials \cite{Dietl:2001_b}. Furthermore, the direct influence of the magnetic field on the hole spectrum was taken into account. Carrier-carrier spin correlation was described by introducing a Fermi-liquid-like parameter $A_F$ \cite{Dietl:1997_a,Haury:1997_a,Jungwirth:1999_a}, which enlarges the Pauli susceptibility of the hole liquid. No disorder effects were taken into account on the ground that their influence on thermodynamic properties is relatively weak except for strongly localized regime. Having the hole energies, the free energy density ${\cal{F}}_c[\vec{M}]$ was evaluated according to the procedure suitable for Fermi liquids of arbitrary degeneracy. By minimizing ${\cal{F}}[\vec{M}] = {\cal{F}}_S[\vec{M}] + {\cal{F}}_c[\vec{M}]$ with respect to $\vec{M}$ at given $T$, $H$, and hole concentration $p$, Mn spin magnetization \index{magnetization} $M(T,H)$ was obtained as a solution of the mean-field equation,
\begin{equation}
\vec{M}(T,H) = x_{eff}N_og\mu_BS\mbox{B}_S[g\mu_B(-\partial {\cal{F}}_c[\vec{M}]/\partial \vec{M} + \vec{H})/k_B(T+T_{AF})],
\end{equation}
where peculiarities of the valence band structure, such as the presence of various hole subbands, anisotropy, and spin-orbit \index{spin-orbit interaction} coupling, are hidden in $F_c[\vec{M}]$. Near the Curie temperature \index{Curie temperature} $T_C$ and at $H = 0$, where $M$ is small, we expect ${\cal{F}}_c[M] - {\cal{F}}_c[0] \sim -M^2$. It is convenient to parameterize this dependence by a generalized carrier spin susceptibility \index{spin susceptibility} $\tilde{\chi}_c$, which is related to the magnetic susceptibility of the carrier liquid according to $\tilde{\chi}_c = A_F(g*\mu_B)^2\chi_c$. In terms of $\tilde{\chi}_c$,
\begin{equation}
{\cal{F}}_c[M] = {\cal{F}}_c[0] - A_F \tilde{\chi_c}\beta^2M^2/2(g\mu_B)^2.
\end{equation}
By expanding B$_S(M)$ for small $M$ one arrives to the mean-field formula for $T_C = T_F - T_{AF}$, where $T_F$ is
given by
\begin{equation}
T_F = x_{eff}N_oS(S+1)A_F\tilde{\chi}_c(T_C)\beta^2/3k_B.
\end{equation}

For a strongly degenerate carrier liquid $|\epsilon_F|/k_BT \gg 1$, as well as neglecting the spin-orbit \index{spin-orbit interaction} interaction  $\tilde{\chi}_c = \rho/4$, where $\rho$ is the total density-of-states for intra-band charge excitations, which in the 3D case is given by $\rho = m^*_{DOS}k_F/\pi^2\hbar^2$. In this case and for $A_F = 1$, $T_F$ assumes the well-known form, derived already in 40s in the context of carrier-mediated nuclear ferromagnetism \cite{Frohlich:1940_a}. In general, however,  $\tilde{\chi}_c$ has to be determined numerically by computing ${\cal{F}}_c[M]$ for a given band structure and degeneracy of the carrier liquid. The model can readily be generalized to various dimensions as well as to the case, when $\vec{M}$ is not spatially uniform in the ground state.

The same formalism, in addition to $T_C$ and Mn magnetization \index{magnetization} $M(T,H)$, as discussed above, provides also quantitative information on spin polarization and magnetization \index{magnetization} of the hole liquid \cite{Dietl:2001_b}. Furthermore, it can be exploited to describe chemical trends as well as micromagnetic, transport, and optical properties of ferromagnetic DMS \index{diluted magnetic semiconductors}, the topics discussed in the subsequent sections.
\subsection{Curie Temperature -- Chemical Trends}
Large magnitudes of both density of states and exchange  \index{exchange} integral specific to the valence band make $T_F$ to be much higher in p-type than in n-type materials with a comparable carrier concentration. Accordingly, in agreement with theoretical evaluations \cite{Dietl:1997_a}, no ferromagnetic order was detected above 1~K in n-(Zn,Mn)O:Al, even when the electron concentration exceeded $10^{20}$~cm$^{-3}$ \cite{Andrearczyk:2001_a}. At the same time, theoretical calculations carried out with no adjustable parameters explained satisfactorily the magnitude of $T_C$ in both (Ga,Mn)As \cite{Dietl:2000_a,Jungwirth:2005_b} and p-type (Zn,Mn)Te \cite{Ferrand:2001_a}. Furthermore, theoretical expectations within the p-d Zener model are consistent with chemical trends in $T_C$ values observed experimentally in (Ga,Mn)Sb, (Ga,Mn)P, (In,Mn)As, (In,Mn)Sb, (Ge,Mn), and p-(Zn,Be)Te though effects of hole localization \cite{Dietl:2001_b,Ferrand:2001_a} preclude the appearance of a uniform ferromagnetic order with a univocally defined $T_C$ value in a number of cases. In addition to localization, a competition between long-range ferromagnetic interactions and intrinsic short-range antiferromagnetic interactions \cite{Kepa:2003_a}, as described by $T_{AF} > 0$ and $x_{eff} < x$, may affect the character of magnetic order \cite{Kechrakos:2005_a}. It appears that the effect is more relevant in II-VI DMS \index{diluted magnetic semiconductors} than in III-V DMS \index{diluted magnetic semiconductors} where Mn centers are ionized, so that the enhanced hole density at closely lying Mn pairs may compensate antiferromagnetic interactions \cite{Dietl:2000_a}. In both groups of materials the density of compensating donor defects appear to grow with the Mn concentration \cite{Matsukura:1998_a,Ferrand:2001_a}. In the case of (Ga,Mn)As the defect involved is the Mn interstitial \cite{Yu:2002_a}, which can be driven and passivated at the surface be low temperature annealing \cite{Edmonds:2004_a}.

According to evaluations carried out by the present author and co-workers \cite{Dietl:2000_a} room temperature ferromagnetism could be observed in a weakly compensated (Ga,Mn)As containing at least 10\% of Mn. At the same time, because of stronger p-d hybridization in wide band-gap materials, such as (Ga,Mn)N and (Zn,Mn)O, $T_C> 300$~K is expected already for $x =5$\%, provided that the hole concentration would be sufficiently high. However, it was clear from the beginning \cite{Dietl:2000_a} that the enhancement of the hole binding energy by p-d hybridization as well as a limited solubility of magnetic constituent together with the effect of self-compensation may render the fabrication of high temperature ferromagnetic DMS \index{diluted magnetic semiconductors} challenging. Nevertheless, a number of group has started the growth of relevant systems, the effort stimulated even further by a number of positive results as well as by numerous theoretical papers suggesting, based on \emph{ab initio} computations, that high temperature ferromagnetism is possible in a large variety of DMS \index{diluted magnetic semiconductors} even without band holes. Today, however, a view appears to prevail that the high temperature ferromagnetism, as evidenced by either magnetic, magnetotransport or magnetooptical phenomena, results actually from the presence of precipitates of known or so-far unknown ferromagnetic or ferrimagnetic nanocrystals containing a high density of magnetic ions. At the same time, it becomes more and more clear that the \emph{ab initio} computations in question suffered from improper treatment of correlation and disorder, which led to an overestimation of tendency towards a ferromagnetic order. It seems at the end that, as argued initially \cite{Dietl:1997_a,Dietl:2000_a}, the delocalized or weakly localized holes are necessary to stabilize a long-range ferromagnetic order in tetrahedrally coordinated DMS \index{diluted magnetic semiconductors} with a small concentration of randomly distributed magnetic ions.

\subsection{Micromagnetic properties}
\subsubsection{Magnetic anisotropy}
As the energy of dipole-dipole magnetic interactions depends on the dipole distribution, there exists the so-called shape anisotropy. In particular, for thin films, the difference in energy density corresponding to the perpendicular and in-plane orientation of magnetization \index{magnetization} M is given by
\begin{equation}
E = \mu_oM^2/2,
\end{equation}
which leads to the anisotropy field $\mu_oH_A  = \mu_oM$ of about 60~mT for Ga$_{0.95}$Mn$_{0.05}$As.

Already early studies of the ferromagnetic phase in (In,Mn)As \cite{Munekata:1993_a}  and (Ga,Mn)As \cite{Shen:1997_a} demonstrated the existence of magnetic anisotropy \index{magnetic anisotropy}, whose character and magnitude implied a sizable contribution of a microscopic origin. Magneto-crystalline anisotropy is usually associated with the interaction between spin and orbital degrees of freedom of the magnetic ion d-electrons. According to the model advocated here, these electrons are in the d$^5$ configuration. For such a case the orbital momentum $L = 0$, so that effects stemming from the spin-orbit \index{spin-orbit interaction} coupling are expected to be rather weak. It was, however, been noted that the interaction between the localized spins is mediated by the holes that have a non-zero orbital momentum $l = 1$ \cite{Dietl:2000_a}. An important aspect of the p-d Zener model is that it does take into account the anisotropy of the carrier-mediated exchange  \index{exchange} interaction associated with the spin-orbit \index{spin-orbit interaction} coupling in the host material \cite{Dietl:2000_a,Dietl:2001_b,Abolfath:2001_a}.

\begin{figure} \centering
\includegraphics*[width=90mm]{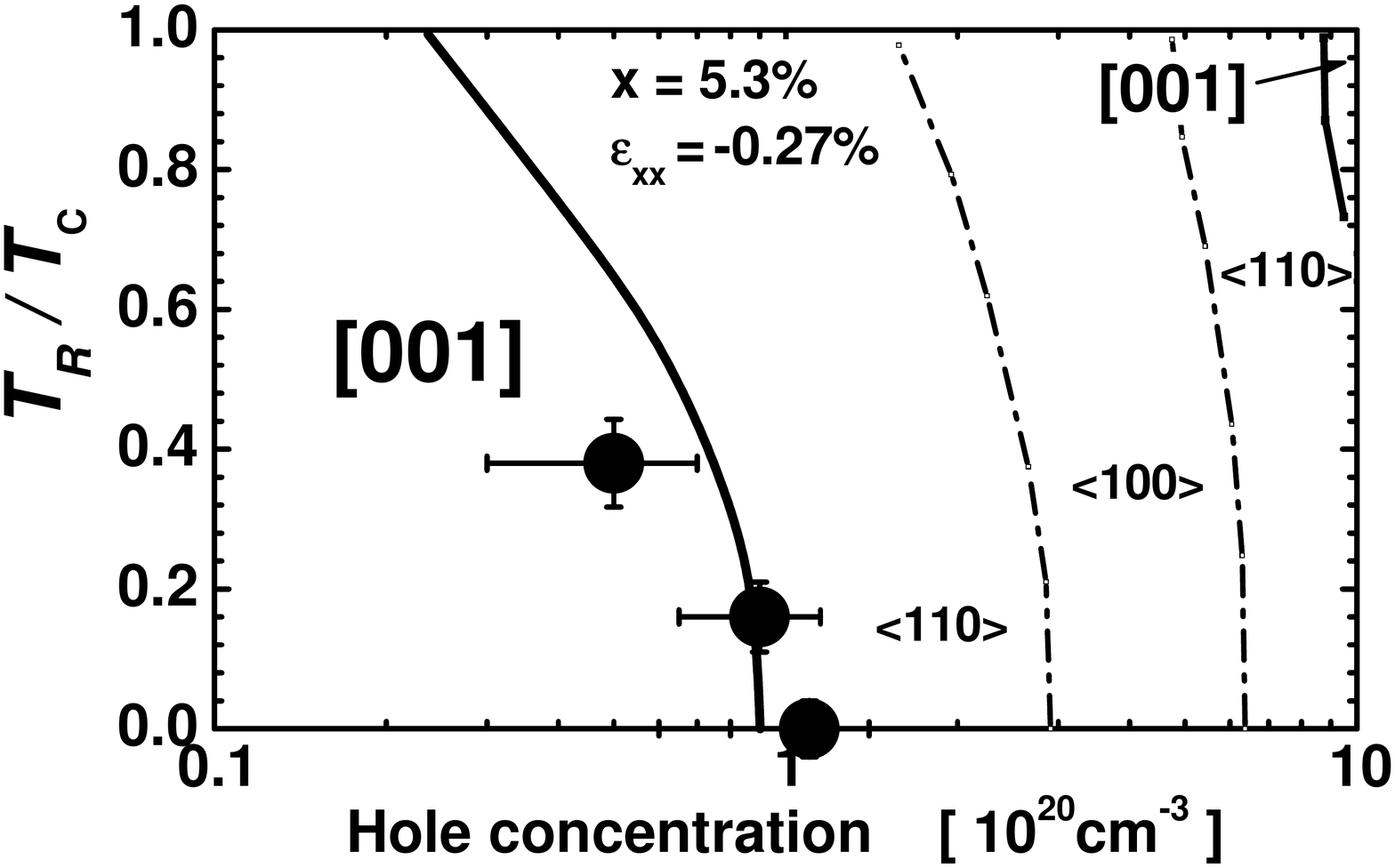}
\caption{Experimental (full points) and computed values (thick lines) of the ratio of the reorientation to Curie temperature \index{Curie temperature} for the transition from perpendicular to in-plane magnetic anisotropy \index{magnetic anisotropy}. Dashed lines mark expected temperatures for the reorientation of the easy axis between $\langle 100\rangle$ and $\langle 110\rangle$ in-plane directions (after Sawicki \emph{et al.} \cite{Sawicki:2004_a}).}
\label{Fig:RT}
\end{figure}

A detail theoretical analysis of anisotropy energies and anisotropy fields in films of (Ga,Mn)As was carried out for a number of experimentally important cases within the p-d Zener model \cite{Dietl:2001_b,Abolfath:2001_a}. In particular, the cubic anisotropy as well as uniaxial anisotropy under biaxial epitaxial strain were examined as a function of the hole concentration $p$.  Both shape and magneto-crystalline anisotropies were taken into account. The perpendicular and in-plane orientation of the easy axis is expected for the compressive and tensile strain, respectively, provided that the hole concentration is sufficiently small. However, according to theory, a reorientation of the easy axis direction is expected at higher hole concentrations. Furthermore, in a certain concentration range the character of magnetic anisotropy \index{magnetic anisotropy} is computed to depend on the magnitude of spontaneous magnetization \index{magnetization}, that is on the temperature. The computed phase diagram for the reorientation transition compared to the experimental results for a film is shown in Fig.~\ref{Fig:RT}. In view that theory is developed with no adjustable parameters the agreement between experimental and computed concentrations and temperature corresponding to the reorientation transition is very good. Furthermore, the computed magnitudes of the anisotropy field $H_u$ \cite{Dietl:2001_b} are consistent with the available findings for both compressive and tensile strain.

According to the discussion above, the easy axis assumes the in-plane orientation for typical carrier concentrations in the most thoroughly studied system (Ga,Mn)As/GaAs. In this case the easy axis is expected to switch between $\langle 100\rangle$ and $\langle 110\rangle$ in-plane cubic directions as a function of $p$ \cite{Dietl:2001_b,Abolfath:2001_a}. Surprisingly, however, only the $\langle 100\rangle$ biaxial magnetic symmetry has so-far been observed in films of (Ga,Mn)As/GaAs at low temperatures. Nevertheless, the corresponding in-plane anisotropy field assumes the expected magnitude, of the order of 0.1~T, which is typically much smaller than that corresponding to the strain-induced energy of magnetic anisotropy \index{magnetic anisotropy}. It is possible that anisotropy of the hole magnetic moment, neglected in the theoretical calculations \cite{Dietl:2001_b,Abolfath:2001_a}, stabilizes the $\langle 100\rangle$ orientation of the easy axis.

In addition to the cubic in-plane anisotropy, the accumulated data for both (Ga,Mn)As/GaAs and (In,Mn)As/(In,Al)As point to a non-equivalence of [110] and [-110] directions, which leads to the in-plane uniaxial magnetic anisotropy \index{magnetic anisotropy}. Such a uniaxial anisotropy is not expected for D$_{2d}$ symmetry of a T$_d$ crystal under epitaxial strain \cite{Sawicki:2005_a,Wang:2005_e}. Furthermore, the magnitude of the corresponding anisotropy field appears to be independent of the film thickness \cite{Welp:2004_a}, which points to a puzzling symmetry breaking in the film body.

\subsubsection{Magnetic stiffness and domain structure}

Another important characteristics of any ferromagnetic system is magnetic stiffness $A$, which describes the energy penalty associated with the local twisting of the direction of magnetization \index{magnetization}. Remarkably, $A$ determines the magnitude and character of thermodynamic fluctuations of magnetization \index{magnetization}, the spectrum of spin excitations as well as the width and energy of domain walls. An important result is that the magnetic stiffness computed within the $6\times 6$ Luttinger model is almost by a factor of 10 greater than that expected for a simple spin degenerate band with the heave-hole band-edge mass \cite{Konig:2001_a}. This enhancement, which stabilizes strongly the spatially uniform spin ordering, stems presumably from p-like symmetry of the valence band wave functions, as for such a case the carrier susceptibility (the Lindhard function) decreases strongly with $q$ \cite{Szymanska:1978_a}.

\begin{figure} \centering
\includegraphics*[width=90mm]{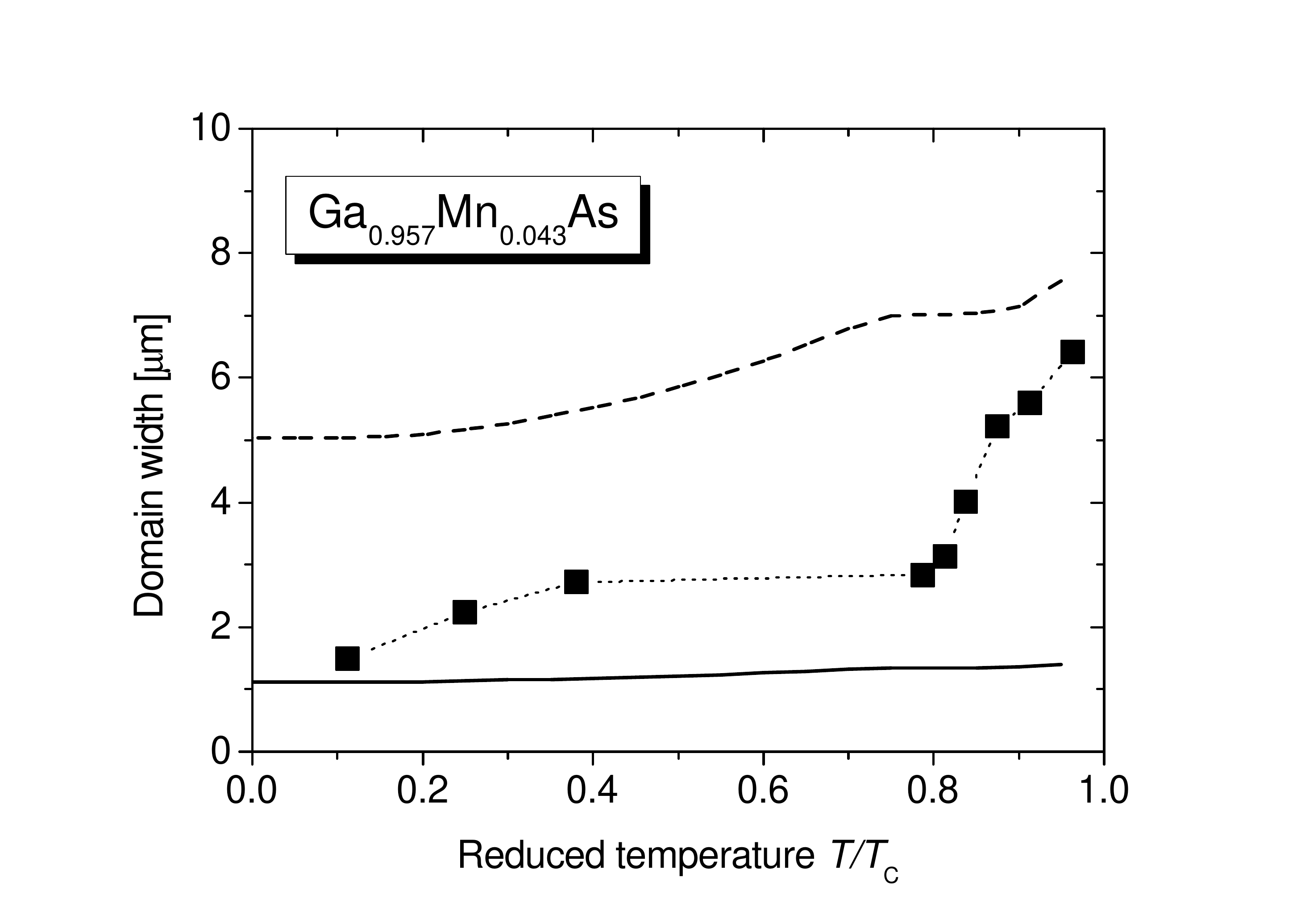}
\caption{Temperature dependence of the width of domain stripes as measured by Shono \emph{et al.} \cite{Shono:2000_a} for the Ga$_{0.957}$Mn$_{0.043}$As film with the easy axis along the growth direction (full squares). Computed domain width is shown by the solid line (after Dietl \emph{et al.} \cite{Dietl:2001_c}).}
\label{Fig:domains}
\end{figure}

The structure of magnetic domains in (Ga,Mn)As under tensile strain has been determined by micro-Hall probe imaging \cite{Shono:2000_a}. The regions with magnetization \index{magnetization} oriented along the [001] and [00-1] easy axis form alternating stripes extending in the [110] direction. As shown in Fig.~\ref{Fig:domains}, the experimentally determined stripe width is $W = 1.5$~$\mu$m at 5~K for 0.2~$\mu$m film of Ga$_{0.957}$Mn$_{0.043}$As on Ga$_{0.84}$In$_{0.16}$As, for which tensile strain of $\epsilon_{xx} = 0.9$\% is expected. According to micromagnetic theory, $W$ is determined by the ratio of the domain wall energy to the stray field energy. As shown in Fig.~\ref{Fig:domains}, the computed value with no adjustable parameters $W = 1.1$~$\mu$m \cite{Dietl:2001_c} compares favorably with the experimental finding, $W = 1.5$~$\mu$m at low temperatures. However, the model predicts much weaker temperature dependence of $W$ than that observed experimentally, which was linked \cite{Dietl:2001_c} to critical fluctuations, disregarded in the mean-field approach.

\subsection{Optical properties}

\subsubsection{Magnetic circular dichroism}

\begin{figure} \centering
\includegraphics[width=90mm]{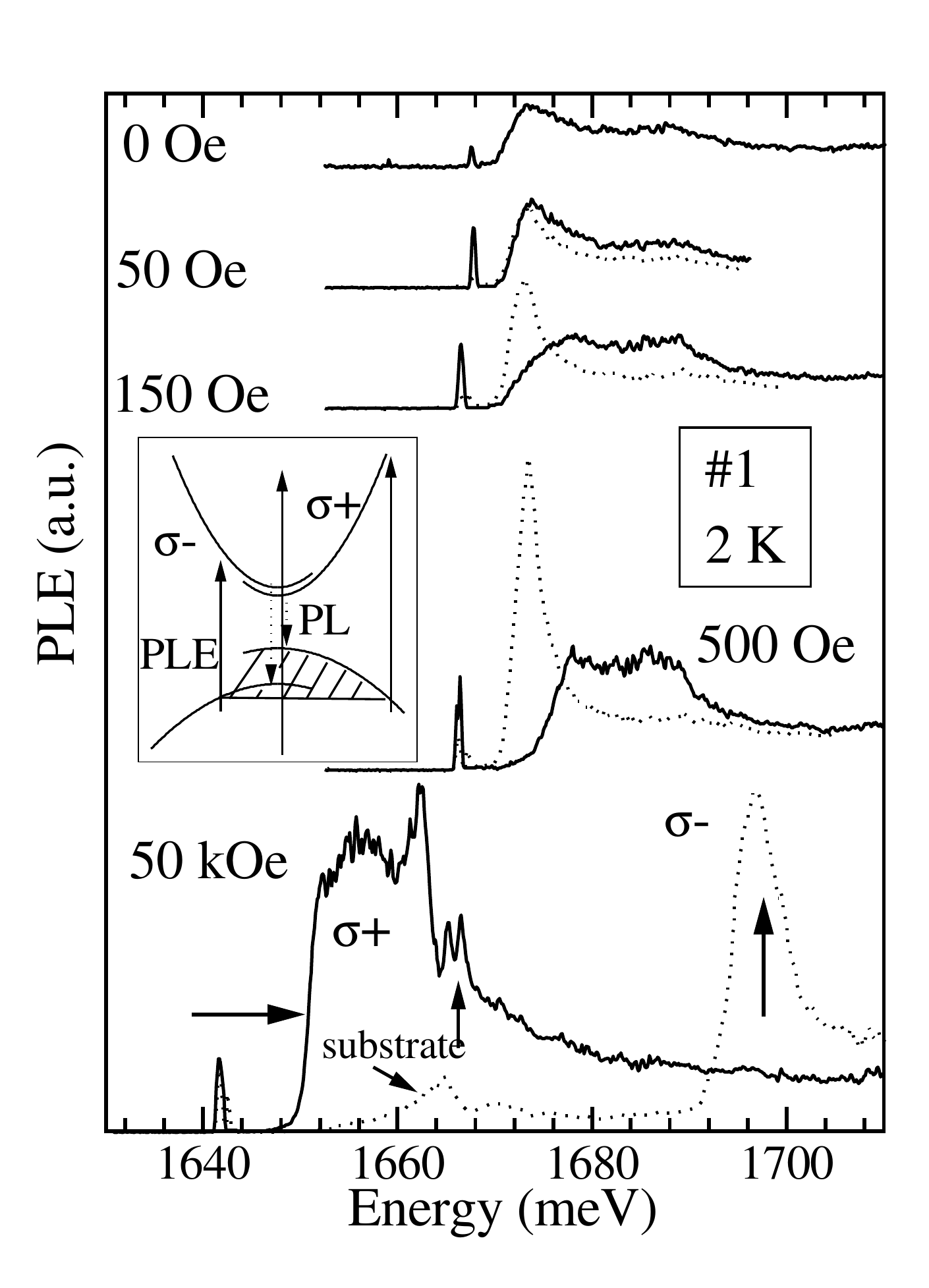}
\caption{Photoluminescence \index{luminescence} excitation spectra (PLE), that is the
photoluminescence \index{luminescence} (PL) intensity as a function of the excitation
photon energy intensity, for $\sigma^+$ (solid lines) and
$\sigma^-$ (dotted lines) circular polarizations at selected values
of the magnetic field in a modulation-doped p-type quantum well of
Cd$_{0.976}$Mn$_{0.024}$Te at 2 K. The photoluminescence \index{luminescence} was
collected in $\sigma^+$ polarization at energies marked by the
narrowest features. The sharp maximum (vertical arrow) and
step-like form (horizontal arrow) correspond to quasi-free exciton
and transitions starting at the Fermi level, respectively. Note
reverse ordering of transition energies at $\sigma^+$ and
$\sigma^-$  for PL and PLE (the latter is equivalent to optical
absorption). The band arrangement at 150~Oe is sketched in the
inset (after Haury \emph{et al.} \cite{Haury:1997_a}).}
\label{Fig:3_SST}
\end{figure}

Within the Zener model, the strength of the ferromagnetic spin-spin
interaction is controlled by the $k\cdot p$ parameters of the host
semiconductor and by the magnitude of the spin-dependent coupling
between the effective mass carriers and localized spins. In the
case of II-VI DMS \index{diluted magnetic semiconductors}, detailed information on the exchange  \index{exchange}-induced
spin-splitting of the bands, and thus on the coupling between the
effective mass electrons and the localized spins has been obtained
from magnetooptical studies \cite{Dietl:1994_a}. A similar work
on (Ga,Mn)As \cite{Ando:1998_a,Szczytko:1999_a,Beschoten:1999_a} led to a number of
surprises. The most striking was the opposite order of the
absorption edges corresponding to the two circular photon
polarizations in (Ga,Mn)As comparing to II-VI materials. This
behavior of circular magnetic dichroism (MCD \index{magnetic circular dichroism}) suggested the
opposite order of the exchange  \index{exchange}-split spin subbands, and thus a
different origin of the sp-d interaction in these two families of
DMS \index{diluted magnetic semiconductors}. A new light on the issue was shed by studies of
photoluminescence \index{luminescence} (PL) and its excitation spectra (PLE) in p-type
(Cd,Mn)Te quantum wells \cite{Haury:1997_a}.  As shown schematically in
Fig.~\ref{Fig:3_SST}, the reversal of the order of PLE edges corresponding to the
two circular polarizations results from the Moss-Burstein effect,
that is from the shifts of the absorption edges associated with the
empty portion of the valence subbands in the p-type material.

\begin{figure} \centering
\includegraphics*[width=90mm]{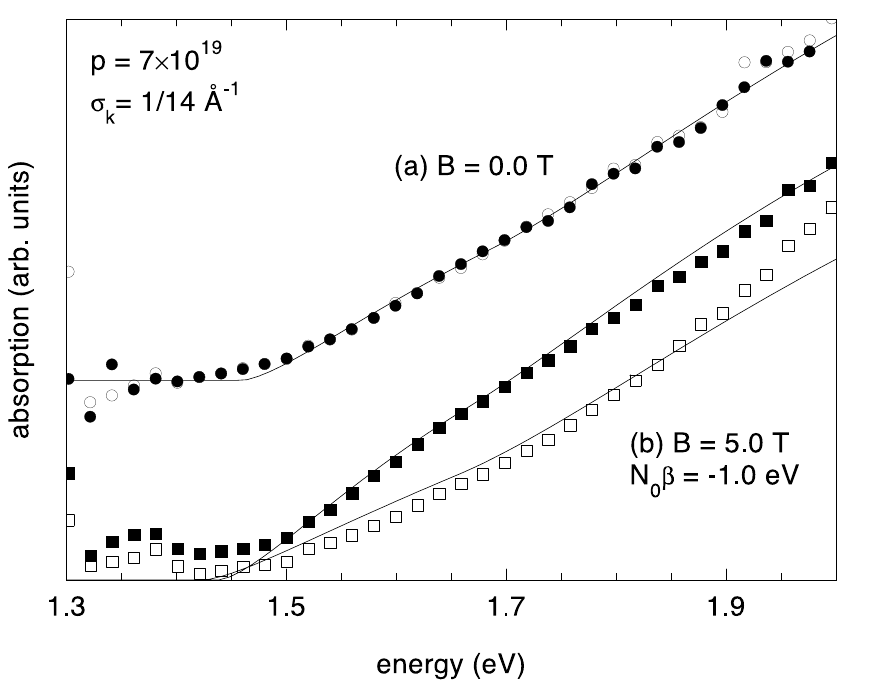}
\caption{Transmission of Ga$_{0.968}$Mn$_{0.032}$As film for two
circular light polarizations in the Faraday configuration in the
absence of the magnetic field (data shifted up for clarity) and in
5~T at 2~K (points) \cite{Szczytko:1999_a}. Solid lines are calculated for
the hole concentration $p=7\times 10^{19}$ cm$^{-3}$, exchange  \index{exchange}
energy $N_0\beta=-1$~eV, and allowing for scattering-induced
breaking of the $k$ selection rules \cite{Szczytko:2001_b}.}
\label{Fig:4_SST}
\end{figure}

The above
model was subsequently applied to interpret the
magnetoabsorption data for metallic (Ga,Mn)As \cite{Szczytko:1999_a,Dietl:2001_b}. More
recently, the theory was extended by taking into account the effect
of scattering-induced mixing of $k$ states \cite{Szczytko:2001_b}. As shown
in Fig.~\ref{Fig:4_SST}, this approach explains the slop of the absorption edge
as well as its field-induced splitting assuming the value of the
p-d exchange  \index{exchange} energy $\beta N_0 = -1$~eV.

Recently, the formalisms suitable for description of either interband \cite{Dietl:2001_b} or intraband \cite{Sinova:2002_a} optical absorption were combined \cite{Hankiewicz:2004_a} in order to examine theoretically optical (dynamic) conductivity in the whole spectral range up to 2~eV.  Furthermore a possible presence of optical absorption involving defect states was taken into account. In this way, the most general quantitative theory of optical and magnetoptical effects in magnetic semiconductors available to date was worked out. A good quantitative description of experimental data \cite{Nagai:2001_a,Singley:2003_a} was obtained verifying the model. However, some discrepancies in the low photon energy range were detected, which confirmed the presence of quantum localization effects. At the same time, a disagreement in the high energy region pointed to the onset of intra-d band transitions. The Faraday and Kerr rotations were also computed showing a large magnitude and a complex spectral dependence in the virtually whole studied photon energy range up to 2 eV, which suggests a suitability of this material family for magnetooptical applications.

\subsection{Charge transport phenomena}

\subsubsection{Hall effect in ferromagnetic semiconductors - theory}

The assessment of magnetic characteristics by means of
magnetotransport studies is of particular importance in the case of
thin films of diluted magnets, in which the magnitude of the total
magnetic moment is typically small. For this reason, recent years
have witnessed a renewed interest in the nature of the anomalous
Hall effect (AHE),
which--if understood theoretically--can serve to determine the
magnitude of magnetization \index{magnetization}. Also magnetoresistance, to be discussed
later on,
 provides information on the magnetism and on the
interplay between electronic and magnetic degrees of freedom.

The Hall resistance $R_{Hall}\equiv \rho_{yx}/d$ of a film of the
thickness $d$ is empirically known to be a sum of ordinary and
anomalous Hall terms in magnetic materials \cite{Chien:1980_a},
\begin{equation}
R_{Hall} = R_0\mu_oH/d + R_S\mu_oM/d.
\label{Eq_1_LH}
\end{equation}
Here, $R_0$ and $R_S$ are the ordinary and anomalous Hall
coefficients, respectively ($R_0> 0$ for the holes), and $M(T,H)$
is the component of the magnetization \index{magnetization} vector perpendicular to the
sample surface. While the ordinary Hall effect serves to determine
the carrier density, the anomalous Hall effect (known also as the
extraordinary Hall effect) provides valuable information on
magnetic properties of thin films. The coefficient $R_S$ is usually
assumed to be proportional to $R_{sheet}^{\alpha}$, where
$R_{sheet}(T,H)$ is the sheet resistance and the exponent $\alpha$
depends on the mechanisms accounting for the AHE.

If the effect of stray magnetic fields produced by localized
magnetic moments were been dominating, $R_S$ would scale with
magnetization \index{magnetization} $M$ but would be
rather proportional to $R_0$ than to $R_{sheet}$. There is
no demagnetization \index{magnetization} effect in the magnetic field perpendicular to
the surface of a uniformly magnetized film, $B=\mu_oH$. However, this is no longer the case
in the presence of magnetic precipitates, whose stray fields and AHE
may produce an apparent magnetization \index{magnetization}-dependent contribution the host Hall resistance.

When effects of stray fields can be disregarded, spin-orbit \index{spin-orbit interaction} interactions control
totally $R_S$. In such a situation $\alpha$ is either 1 or 2
depending on the origin of the effect: the skew-scattering (extrinsic)
mechanism, for which the Hall conductivity is proportional to
momentum relaxation time $\tau$, results in $\alpha \approx 1$
\cite{Chien:1980_a}. From the theory point of
view particularly interesting is the intrinsic mechanism for
the Hall conductivity
$\sigma_{AH}=R_SM/(R_{sheet}d)^2]$ does not depend explicitly on
scattering efficiency but only on the band structure parameters
\cite{Luttinger:1958_a,Jungwirth:2002_a,Jungwirth:2006_a}.

For both extrinsic and intrinsic mechanisms, the overall
magnitude of the anomalous Hall resistance depends on the strength
of the spin-orbit \index{spin-orbit interaction} interaction and spin polarization of the carriers
at the Fermi surface. Accordingly, at given magnetization \index{magnetization} $M$, the
effect is expected to be much stronger for the holes than for the
electrons in tetrahedrally coordinated semiconductors. For the
carrier-mediated ferromagnetism, the latter is proportional to the
exchange  \index{exchange} coupling of the carriers to the spins, and varies -- not
necessarily linearly -- with the magnitude of spin magnetization \index{magnetization}
$M$. Additionally, the skew-scattering contribution depends on the
asymmetry of scattering rates for particular spin subbands, an
effect which can depend on $M$ in a highly nontrivial way.
Importantly, the sign of either of the two contributions can be
positive or negative depending on a subtle interplay between the
orientations of orbital and spin momenta as well as on the
character (repulsive vs. attractive) of scattering potentials.

Recently, Jungwirth {\it et al.} \cite{Jungwirth:2002_a} developed a theory
of the intrinsic AHE in p-type zinc-blende magnetic semiconductors, and
presented numerical results for the case of (Ga,Mn)As, (In,Mn)As,
and (Al,Mn)As. The derived formula for $\sigma_{AH}$ corresponds
to that given earlier \cite{Luttinger:1958_a,Nozieres:1973_a,Chazalviel:1975_a} in the weak scattering limit.
The intrinsic AHE can also
be regarded as a zero-frequency limit of $\sigma_{xy}(\omega)$, where
$\sigma(\omega)$ is the dynamic (optical) conductivity tensor, related directly
the Kerr effect, widely studied in experimentally and theoretically
in ferromagnetic metals \cite{Oppener:2001_a}.
For the hole concentration
$p$ such that the Fermi energy is much smaller than the spin-orbit \index{spin-orbit interaction}
splitting $\Delta_o$ but larger than the exchange  \index{exchange} splitting $h$
between the majority $j_z=-3/2$ and minority $j_z=+3/2$ bands at
$k=0$, $\Delta_o \gg |\epsilon_F| \gg h$, Jungwirth {\it et
al.} \cite{Jungwirth:2002_a} predict within the $4\times 4$ spherical
Luttinger model
\begin{equation}
\sigma_{AH}^{in}=e^2hm_{hh}/[4\pi^2\hbar^3(3\pi p)^{1/3}].
\label{Eq_4_LH}
\end{equation}
Here the heavy hole mass $m_{hh}$ is assumed to be much larger than
the light hole mass $m_{lh}$, whereas $\sigma_{AH}^{in}$ becomes by
the factor of $2^{4/3}$ greater in the opposite limit
$m_{hh}=m_{lh}$. In the range $h \ll |\epsilon_F| \ll \Delta_o$ the
determined value of $\sigma_{AH}^{in}$ is positive, that is the
coefficients of the normal and anomalous Hall effects are expected
to have the same sign. However, if the Fermi level were approached
the split-off $\Gamma_7$ band, a change of sign would occur.

A formula for $\sigma_{AH}^{in}$ was also derived \cite{Dietl:2003_c} from
Eq.~4 of Jungwirth {\it et al.} \cite{Jungwirth:2002_a}), employing the known
form of the heavy hole Bloch wave functions $u_{\vec{k},j_z}$
\cite{Szymanska:1978_a}.  Neglecting a small effect of the spin splitting on
the heavy hole wave functions, $\sigma_{AH}^{in}$ was found to be
given by the right hand side of Eq.~\ref{Eq_2_LH} multiplied by the factor
$(16/9)\ln2-1/6\approx 1.066$ \cite{Dietl:2003_c}.

In order to evaluate the ratio of intrinsic and
skew-scattering mechanisms, the general theory of the AHE effect in semiconductors
\cite{Luttinger:1958_a,Nozieres:1973_a,Chazalviel:1975_a,Jungwirth:2002_a} was applied \cite{Dietl:2003_a}.
Assuming that scattering  by ionized impurities dominates,
this ratio is then given by \cite{Leroux-Hugon:1972_a,Chazalviel:1974_a,Chazalviel:1975_a},
\begin{equation}
\frac{\sigma_{AH}^{in}}{\sigma_{AH}^{ss}}= \pm f(\xi)(N_A+N_D)
/(pr_sk_F\ell).
\label{Eq_2_LH}
\end{equation}

Here, $f(\xi) \approx 10$ is a function that depends weakly on the
screening dimensionless parameter $\xi$; $(N_A+N_D)/p$ is the ratio
of the ionized impurity and carrier concentrations; $r_s$ is the
average distance between the carriers in the units of the effective
Bohr radius, and $\ell$ is the mean free path. Similarly, for
spin-independent scattering by short range potentials, $V(
\vec{r}) = V\delta(\vec{r} - \vec{r}_i)$ \cite{Nozieres:1973_a}
was applied \cite{Dietl:2003_c}.
Assuming that scattering  by ionized impurities is negligible,
\begin{equation}
\frac{\sigma_{AH}^{in}}{\sigma_{AH}^{ss}}= - 3/[\pi
V\rho(\varepsilon_F)k_F\ell],
\label{Eq_3_LH}
\end{equation}
where $\rho(\varepsilon_F)$ is the density of states at the Fermi
level. Of course, the overall sign depends on the sign of the
scattering potential $V$.

In order to find out which of the two AHE mechanisms operates
predominantly in p-type tetrahedrally coordinated ferromagnetic
semiconductors, we note that scattering by ionized impurities
appears to dominate in these heavily doped and compensated
materials. This scattering mechanism, together with alloy and spin
disorder scattering, limits presumably the hole mobility and leads
ultimately to the metal-to-insulator transition (MIT). Since at the
MIT $r_s\approx 2$ and $k_F\ell \approx 1$ one expects from Eq.~\ref{Eq_3_LH}
that as long as the holes remain close to the localization boundary
the intrinsic mechanism accounts for the AHE. It would be
interesting to know how quantum localization corrections affect the
anomalous Hall conductivity as well as how to extend theory towards
the insulator side of the MIT. A work in this direction was
reported \cite{Dugaev:2001_a}.

Obviously, the presence of the AHE makes a meaningful determination
of the carrier type and density difficult in ferromagnetic
semiconductors. Usually, the ordinary Hall effect dominates only in
rather high magnetic fields or at temperatures several times larger
than $T_C$. It appears, therefore, that a careful experimental and
theoretical examination of the resistivity tensor in wide field and
temperature ranges is necessary to separate characteristics of the
spin and carrier subsystems.

\subsubsection{Comparison between theoretical and experimental results}

As mentioned above, because of the dominance of the anomalous Hall
term in wide temperature and field ranges, it is not
straightforward to determine the carrier type and concentration in
ferromagnetic semiconductors. Only at low temperatures and under
very high fields, the anomalous Hall term saturates, so that the
ordinary Hall coefficient can be determined from the remaining
linear change of the Hall resistance in the magnetic field. Note
that although magnetization \index{magnetization} saturates in relatively low magnetic
fields, the negative MR usually persists, and generates the field
dependence of the anomalous Hall coefficient.

\begin{figure} \centering
\includegraphics*[width=90mm]{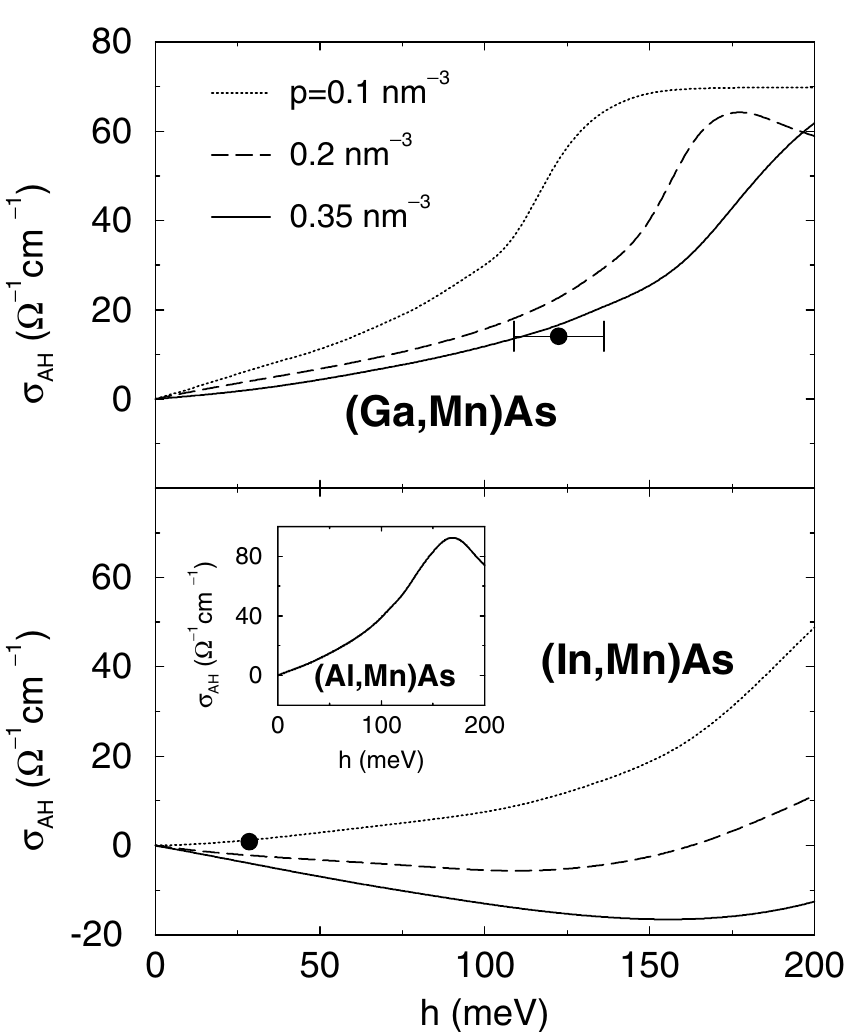}
\caption{Full numerical simulations of the anomalous Hall
conductivity $\sigma_{AH}$ for GaAs host with hole densities $p =
10^{20}$, (dotted lines), $2\times 10^{20}$ (dashed lines), and
$3.5\times 10^{20}$ cm$^{-3}$ (solid lines).  Filled circle
represents measured Hall conductivity (Fig.~2). The saturation
mean-field value of the splitting $h$ between $\Gamma_8$ heavy hole
subbands  was estimated from nominal sample parameters. Horizontal
error bar corresponds to the experimental uncertainty of the p$-$d
exchange  \index{exchange} integral. Experimental hole density in the (Ga,Mn)As
sample is $3.5\times 10^{20}$ cm$^{-3}$ (after Jungwirth \emph{et al.}
\cite{Jungwirth:2002_a}).}
\label{Fig_4_LH}
\end{figure}

Magnetotransport data collected for (Ga,Mn)As in a wide
temperature and field ranges \cite{Matsukura:1998_a,Omiya:2000_a}
were exploited to test the theory of the AHE \cite{Jungwirth:2002_a}. The results of
such a comparison are shown in Fig.~\ref{Fig_4_LH}. There is a
good agreement between the theoretical and experimental magnitude
of the Hall conductivity. Importantly, no significant contribution
from skew scattering is expected for the (Ga,Mn)As sample in
question \cite{Omiya:2000_a}, for which $(N_A+N_D)/p \approx
5$, $r_s \approx 1.1$, and $k_F\ell \approx 0.8$, so that
$\sigma_{AH}^{in}/\sigma_{AH}^{ss}\approx 57$.

\begin{figure} \centering
\includegraphics*[width=90mm]{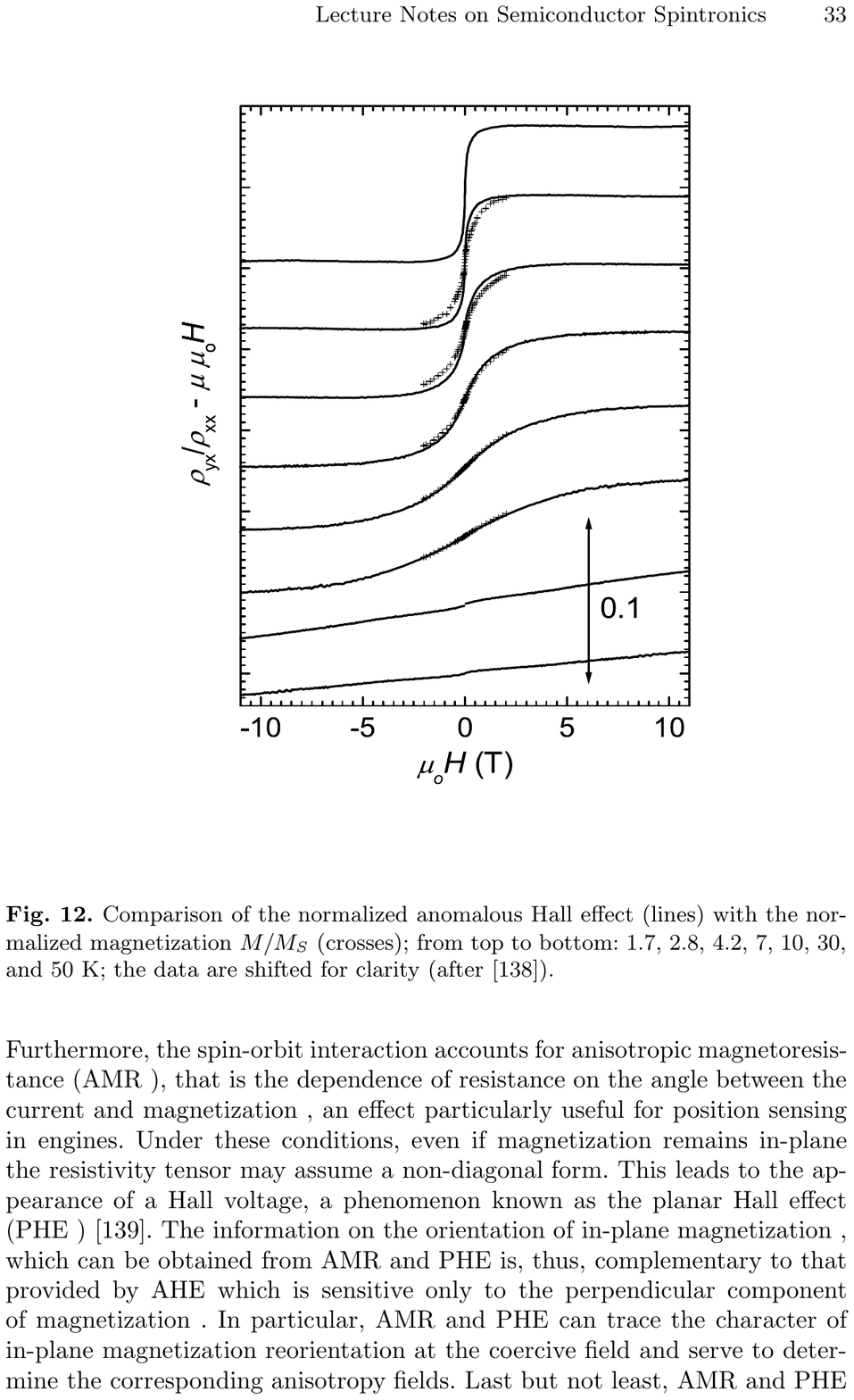}
\caption{Comparison of the normalized
anomalous Hall effect (lines) with the normalized magnetization \index{magnetization}
$M/M_S$ (crosses); from top to bottom: 1.7, 2.8, 4.2, 7, 10, 30,
and 50~K; the data are shifted for clarity (after \cite{Ferrand:2000_a}).}
\label{Fig_6_LH}
\end{figure}

Another material for which various contributions
to Hall resistance were analyzed is Zn$_{0.981}$Mn$_{0.019}$Te:N
containing $1.2\times 10^{20}$ holes per cm$^3$ \cite{Dietl:2003_c}.
In Fig.~\ref{Fig_6_LH}, $\rho_{yx}/\rho_{xx} - \mu B$, i.e., the spin dependent
Hall angle, is compared to the magnetization \index{magnetization} measured in a
vibrating sample magnetometer \cite{Ferrand:2000_a} for this film. The normal Hall angle
$\mu B =\mu\mu_oH$ was subtracted assuming a constant hole mobility
$\mu$~i.e., assigning the conductivity changes entirely to
variations in the hole concentration. This assumption is not
crucial for the present highly doped sample, but it proves to be
less satisfactory for the less doped samples. As shown in Fig.~\ref{Fig_6_LH}, a
reasonable agreement is found by taking,
\begin{equation}
\rho_{yx}/\rho_{xx} = \mu B + \Theta M/M_S,
\label{Eq_5_LH}
\end{equation}
where $M_S$ is the saturation value of magnetization \index{magnetization} and $\Theta =
0.04$ is the adjustable parameter.  For the sample in question, the
maximum value of hole polarization, $(p^{up}-p^{down})/(p^{up}
+p^{down})$, has been estimated to be of the order of 10\%
\cite{Ferrand:2000_a}.

Here, similarly to the case of (Ga,Mn)As, the sign and
magnitude of the anomalous Hall coefficient indicated that the intrinsic
mechanism is involved. The value of
 $\Theta$ was evaluated theoretically from Eq.~\ref{Eq_4_LH} by adopting parameters
suitable for the sample in question, $m_{hh} = 0.6m_o$, $\rho_{xx}
= 5\times 10^{-3}$~$\Omega$cm and the saturation value of the
splitting $h=41$~meV. This leads to $\sigma_{AH}^{in}
=13.1$~($\Omega$cm)$^{-1}$ and $\Theta^{in} = 0.065$ \cite{Dietl:2003_c}, in a
reasonable agreement with the experimental value $\Theta = 0.04$.
Since a contribution from the light hole band will enhance the
theoretical value, it was concluded \cite{Dietl:2003_c} that the present theory describes
the anomalous hole effect within the factor of about two.

It is important to note that there exist several reasons causing
that the Hall effect and direct magnetometry can provide different
information on magnetization \index{magnetization}. Indeed, contrary to the standard
magnetometry, the AHE does not provide information about the
magnetization \index{magnetization} of the whole samples but only about its value in
regions visited by the carriers. Near the metal-insulator boundary,
especially when the compensation is appreciable, the carrier
distribution is highly non-uniform. In the regions visited by the
carriers the ferromagnetic interactions are strong, whereas the
remaining regions may remain paramagnetic. Under such conditions,
magnetotransport and direct magnetic measurements will provide
different magnetization \index{magnetization} values \cite{Dietl:2000_a}. In particular, $M_S$
at $T \rightarrow 0$, as seen by a direct magnetometry, can be much
lower than that expected for a given value of the magnetic ion
concentration. High magnetic fields are then necessary to magnetize
all localized spins. The corresponding field magnitude is expected
to grow with the temperature and strength of antiferromagnetic
interactions that dominate in the absence of the holes.

\subsubsection{Anisotropic magnetoresistance and planar Hall effect}

In cubic materials the conductivity tensor is diagonal in the absence of an external magnetic field. However, non-zero values of strain make the resistance to depend on the orientation of current in respect to crystallographic axes. Furthermore, the spin-orbit \index{spin-orbit interaction} interaction accounts for anisotropic magnetoresistance (AMR  \index{anisotropic magnetoresistance}), that is the dependence of resistance on the angle between the current and magnetization \index{magnetization}, an effect particularly useful for position sensing in engines. Under these conditions, even if magnetization \index{magnetization} remains in-plane the resistivity tensor may assume a non-diagonal form. This leads to the appearance of a Hall voltage, a phenomenon known as  the planar Hall effect (PHE  \index{planar Hall effect}) \cite{Tang:2003_a}. The information on the orientation of in-plane magnetization \index{magnetization}, which can be obtained from AMR  \index{anisotropic magnetoresistance} and PHE  \index{planar Hall effect} is, thus, complementary to that provided by AHE which is sensitive only to the perpendicular component of magnetization \index{magnetization}. In particular, AMR  \index{anisotropic magnetoresistance} and PHE  \index{planar Hall effect} can trace the character of in-plane magnetization \index{magnetization} reorientation at the coercive field and serve to determine the corresponding anisotropy fields. Last but not least, AMR  \index{anisotropic magnetoresistance} and PHE  \index{planar Hall effect} are sensitive probe of spin anisotropy at the Fermi surface associated with the strain and spin-orbit \index{spin-orbit interaction} interaction for non-zero magnetization \index{magnetization}. The corresponding theory of AMR  \index{anisotropic magnetoresistance} was developed by Jungwirth \emph{et al.} \cite{Jungwirth:2002_c} within the Drude-Boltzmann formulation of charge transport in solids.

\begin{figure} \centering
\includegraphics*[width=120mm]{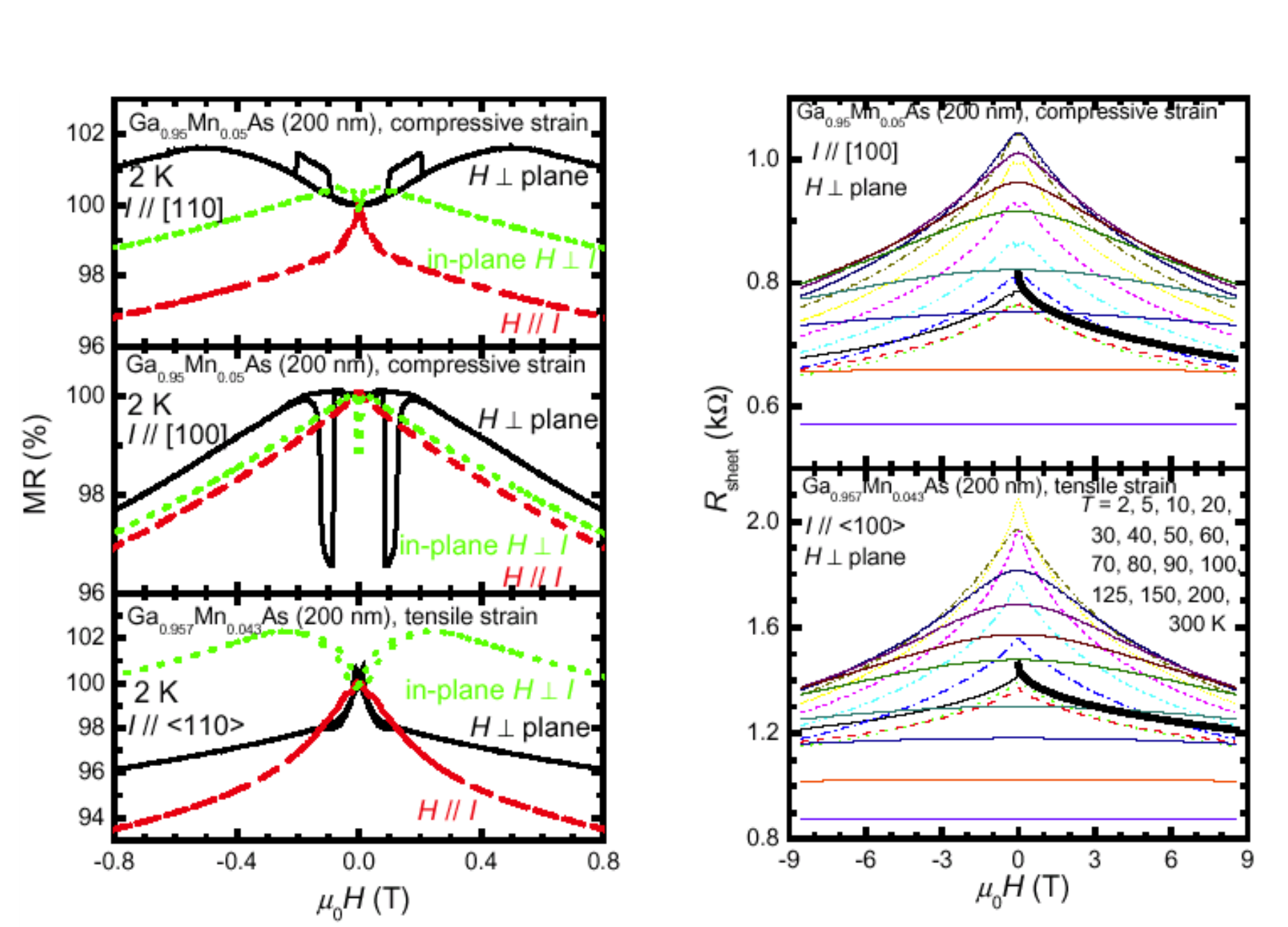}
\caption{Left panel: Field-induced changes in resistance of Ga$_{0.95}$Mn$_{0.05}$As/GaAs (compressive strain) lower panel: current along [110]; middle panel: current along [100]) and of Ga$_{0.957}$Mn$_{0.043}$As/(In,Ga)As  under tensile strain (lower panel, current along $\langle 110\rangle$) for three orientations of the magnetic field in respect to current direction at 2~K. Right panel: field and temperature dependencies of resistance in Ga$_{0.95}$Mn$_{0.05}$As/GaAs (compressive strain, upper panel) and in tensile strained Ga$_{0.957}$Mn$_{0.043}$As/(In,Ga)As (lower panel) for magnetic field perpendicular to the film plane. Starting from up, subsequent curves at $H = 0$ correspond to temperatures in K: 70, 60, 80, 50, 90, 40, 100, 30, 125, 20, 2, 5, 10, 150, 200, 300 (upper panel) and to 50, 60, 40, 70, 30, 80, 90, 20, 100, 2, 10, 5, 125, 150, 200, 300 (lower panel). The thick solid lines superimposed on 2~K data in positive magnetic field side show Kawabata's theory predictions  (after Matsukura \emph{et al.} \cite{Matsukura:2004_a}).}
\label{Fig_3_4_FM}
\end{figure}

To test the theoretical predictions concerning effects of biaxial strain upon AMR  \index{anisotropic magnetoresistance}, (Ga,Mn)As samples under compressive and tensile strain were studied for longitudinal and two perpendicular orientations of the magnetic field in respect to electric current \cite{Matsukura:2004_a}. As show in Fig.~\ref{Fig_3_4_FM}, above 0.5~T, negative magnetoresistance is observed, whose magnitude is virtually independent of experimental configuration. However, the absolute value of resistance $\rho$ in this range depends on the field direction, which is the signature of AMR  \index{anisotropic magnetoresistance}.   These data provide information on processes of the field-induced rotation of magnetization \index{magnetization} for various orientations of the field in respect to crystal and easy axes. In particular, the values of the field corresponding to the resistance maxima are expected to be of the order of the anisotropy field.

\

If only spin-orbit \index{spin-orbit interaction} effects were controlled AMR  \index{anisotropic magnetoresistance}, its magnitude would depend only on the angle between the current and field directions. According to Fig.~\ref{Fig_3_4_FM}, this is not the case since AMR  \index{anisotropic magnetoresistance} depends also on the directions of the field and current in respect to crystal axes. It is convenient to introduce AMR  \index{anisotropic magnetoresistance}$_{op} = [\rho_{xx}(H \parallel x) - \rho_{xx}(H \parallel  y)]/\rho_{xx}(H \parallel  y)$ and AMR  \index{anisotropic magnetoresistance}$_{op} = [\rho_{xx}(H \parallel x) - \rho_{xx}(H \parallel  z)]/\rho_{xx}(H \parallel z)$, where the current and growth directions are denoted by $x$ and $z$, respectively, and $\rho_{xx}$ is the longitudinal resistivity. Importantly, the sign and order of magnitude of AMR  \index{anisotropic magnetoresistance} is consistent with theoretical expectations \cite{Jungwirth:2002_c,Jungwirth:2003_b}. In particular, the predicted difference in sign of AMR  \index{anisotropic magnetoresistance}$_{op} - $AMR  \index{anisotropic magnetoresistance}$_{ip}$ in the case of compressive and tensile strain is corroborated by the data. On the other hand, the dependence of AMR  \index{anisotropic magnetoresistance}$_{op}$ and AMR  \index{anisotropic magnetoresistance}$_{ip}$ at given strain on the current direction appears as challenging. It may result from the lowering of the symmetry of the (Ga,Mn)As films from the expected D$_{2d}$ to C$_{2v}$, as discussed in Sec.~3.

Particularly intriguing are hysteretic resistance jumps observed for samples under compressive strain and for the field pointing along the growth direction. We assign this effect to a large ratio of the anisotropy and coercive fields, which makes that even a rather small misalignment, and thus a minute in-plane field, can  result in magnetization \index{magnetization} switching between in-plane easy directions. These results provide, therefore, information on resistance values in a demagnetized state for the studied current directions.

\subsubsection{Low and high field magnetoresistance}

Apart from AMR  \index{anisotropic magnetoresistance}, there is a number of other effects that can produce a sizable magnetoresistance in magnetic semiconductors, especially in the vicinity of the localization boundary \cite{Dietl:1994_a}, where quantum corrections to Drude-Boltzmann conductivity become important. In particular, carrier diffusion in the molecular field of randomly oriented spin clusters that form above $T_C$ shifts the metal-to-insulator transition towards higher carrier concentrations \cite{Jaroszynski:2005_b}. The resulting temperature dependent localization may lead to a resistance maximum at $T_C$, which will be destroyed by the magnetic field.  This accounts presumably for the field and temperature dependence of resistivity near $T_C$ visible clearly in Fig.~\ref{Fig_3_4_FM}.

However, the negative magnetoresistance hardly saturates even in rather strong magnetic fields, and occurs also at low temperatures, where the spins are fully ordered ferromagnetically according to the Hall effect data. This surprising observation was explained by the present author and co-workers \cite{Dietl:2003_c,Matsukura:2004_a} in terms of weak localization \index{weak localization} orbital magnetoresistance. Indeed, in the regime in question the giant splitting of the valence band makes both spin-disorder and spin-orbit \index{spin-orbit interaction} scattering relatively inefficient. Under such conditions, weak localization \index{weak localization} magnetoresistance can show up at low temperatures, where phase breaking scattering ceases to operate. According to Kawabata \cite{Kawabata:1980_a},

\begin{equation}
 \triangle \rho/\rho \approx -\triangle \sigma/\sigma = -n_ve^2C_o(e/\hbar B)^{1/2}/(2\pi^2\hbar),
\label{Eq_1_FM}
\end{equation}
where $C_o = 0.605$, $\sigma$ is the conductivity, and $1/2 \leq n_v \leq 2$ depending on whether one or all four hole subbands contribute to the charge transport. For the samples under compressive and tensile strain, the above formula gives $\triangle \rho/\rho= -0.13 n_v$ and $-0.25n_v$, respectively at $B = 9$~T. These values are to be compared to experimental data of Fig.~\ref{Fig_3_4_FM}, $\triangle \rho/\rho = -0.09$ and -0.14 at 2~K. The fitting to Eq.~\ref{Eq_1_FM}  reproduces the data at 2~K quite well (thin solid lines in Fig.~\ref{Fig_3_4_FM} and gives $n_v = 1.46$ and 0.82 for the compressive and tensile samples, respectively, as could be expected for ferromagnetic films of (Ga,Mn)As. Since negative magnetoresistance takes over above $B_i \approx 1$~T, we can evaluate a lower limit for the spin-disorder scattering time, $\tau_s = m*/(eB_ik_Fl) = 8$~ps for the hole effective mass $m*=0.7m_o$ and $k_Fl = 0.5$, where $k_F$ the Fermi momentum and $l$ the mean free path.

\subsection{Spin transport phenomena}

To this category belongs a number of effects observed in heterostructures of (Ga,Mn)As, and important for perspective spintronic devices, such as spin injection \index{spin injection} of holes \cite{Ohno:1999_b,Young:2002_a} and electrons in the Zener diode \cite{Kohda:2001_a,Johnston-Halperin:2002_a}, giant magnetoresistance (GMR  \index{giant magnetoresistance}) \cite{Chiba:2000_a}, tunnelling magnetoresistance (TMR  \index{tunnelling magnetoresistance}) \cite{Tanaka:2001_a,Mattana:2003_a,Chiba:2004_a}, tunnelling anisotropic magnetoresistance (TAMR  \index{anisotropic magnetoresistance}) \cite{Ruster:2005_a,Giddings:2005_a}, and domain wall resistance \cite{Tang:2004_c,Chiba:2006_a}.

Since in most semiconductor spin transport devices the relevant length scale is shorter that the phase coherence length, a formulation of theory in terms of the Boltzmann distribution function $f$ is not valid. Recently, theory that combines an empirical tight-binding approach with a
Landauer-B\"{u}ttiker formalism was developed \cite{Dorpe:2005_c,Sankowski:2005_b}.  In contrast to the
standard $kp$ method, this theory describes properly the interfaces and inversion symmetry
breaking as well as the band dispersion in the entire Brillouin zone, so that the essential for the spin-dependent tunnelling Rashba \index{Rashba hamiltonian} and Dresselhaus terms
as well as the tunnelling \emph{via} $\vec{k}$ points away from the zone center are taken
into account. This approach \cite{Dorpe:2005_c,Sankowski:2005_b}, developed with no adjustable parameters, explained experimentally observed large magnitudes of both electron current
spin polarization up to 70\% in the (Ga,Mn)As/n-GaAs Zener diode \cite{Dorpe:2004_a}  and TMR  \index{tunnelling magnetoresistance} of the order of 300\% in a (Ga,Mn)As/GaAs/(Ga,Mn)As trilayer structure \cite{Chiba:2004_a}. Furthermore, theory reproduced a fast decrease of these figures with the device bias as well as it indicated that the magnitude of TAMR  \index{anisotropic magnetoresistance} should not exceed 10\% under usual strain conditions.

\subsection{Methods of Magnetization Manipulation}

\begin{figure} \centering
\includegraphics*[width=100mm]{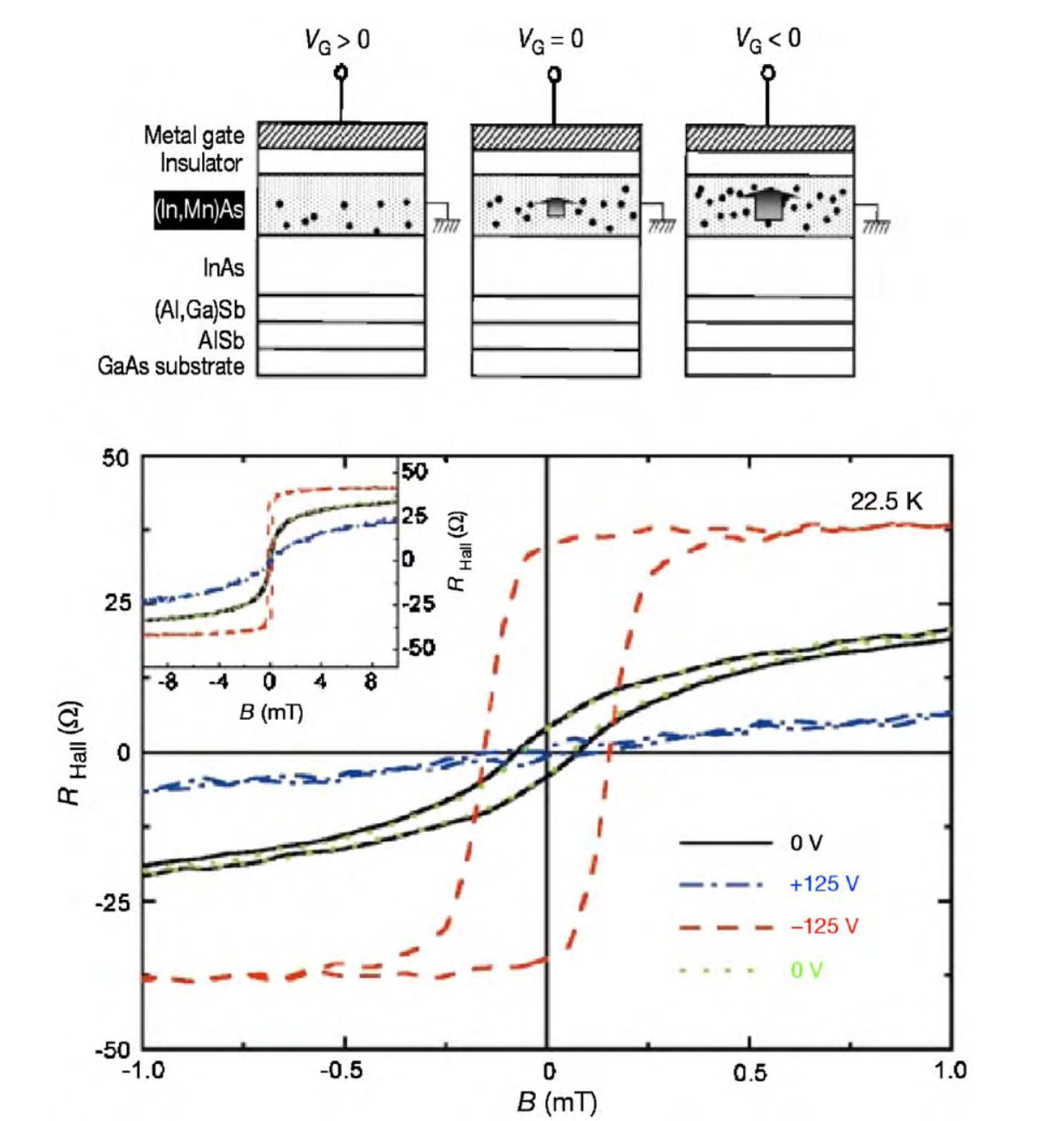}
\caption[angle=270]{Magnetization hysteresis loops determined by measurements of anomalous Hall effect at constant temperature of 22.5~K for various gate voltages in field-effect transistor with (In,Mn)As channel (after Ohno \emph{et al.}~\cite{Ohno:2000_a}).}
\label{FET}
\end{figure}

\begin{figure} \centering
\includegraphics*[angle=270,width=120mm]{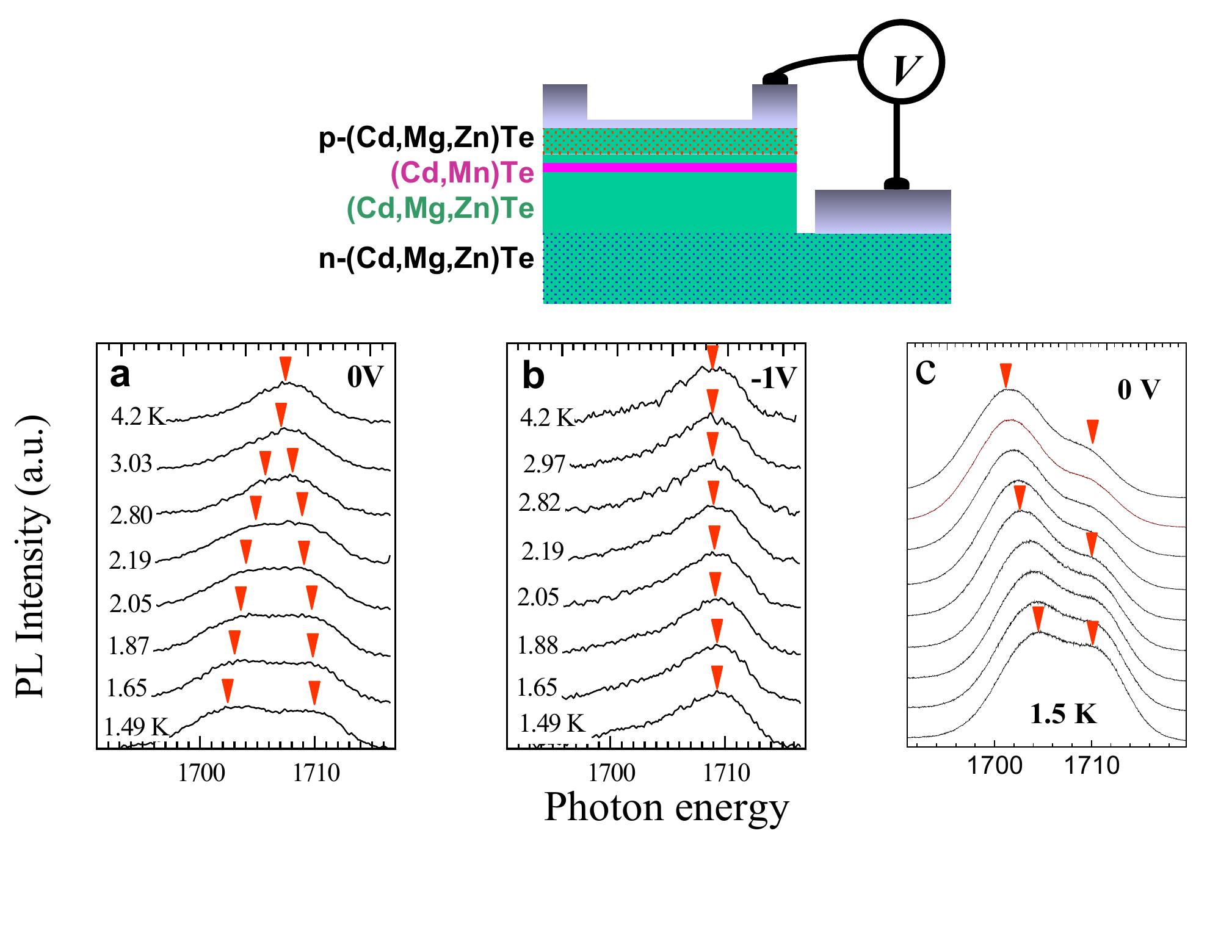}
\caption{Effect of temperature (a), bias voltage (b), and illumination (c) on photoluminescence \index{luminescence} of structure consisting of modulation doped p-(Cd,Mn)Te quantum well and n-type barrier. Zero-field line splitting (marked by  arrows) witnesses the appearance of a ferromagnetic ordering (a) which does not show up if the quantum well is depleted from the holes by reverse bias of p-i-n diode (b). Low-temperature splitting is enhanced by additional illumination by white light (c), which increases  hole concentration in quantum well (after Boukari \emph{et al.}~\cite{Boukari:2002_a}).}
\label{LED}
\end{figure}

Since magnetic properties are controlled by band holes, an appealing possibility is to influence the magnetic ordering isothermally, by light or by the electric field, which affect the carrier concentration in semiconductor structures. Such tuning capabilities of the materials systems in question were put into the evidence in (In,Mn)As/(Al,Ga)Sb \cite{Koshihara:1997_a,Ohno:2000_a} and modulation doped p-(Cd,Mn)Te/(Cd,Mg,Zn)Te \cite{Haury:1997_a,Boukari:2002_a} heterostructures, as depicted in Figs.~\ref{FET} and \ref{LED}. Actually, these findings can be quantitatively interpreted by considering the effect of the electric field or illumination on the hole density under stationary conditions and, therefore, on the Curie temperature \index{Curie temperature} in the relevant magnetic layers. Interestingly, according to experimental findings and theoretical modelling, photocarriers generated in II-VI systems by above barrier illumination destroy ferromagnetic order in the magnetic quantum well residing in an undoped (intrinsic) region of a p-i-p structure \cite{Haury:1997_a,Boukari:2002_a} but they enhance the magnitude of spontaneous magnetization \index{magnetization} in the case of a p-i-n diode \cite{Boukari:2002_a}, as shown in Fig.~\ref{LED}.

Another method of magnetization \index{magnetization} manipulation, suitable for low-power switching of bits in magnetic memories, was invoked by Luc Berger \cite{Berger:1984_a} and John Slonczewski \cite{Slonczewski:1996_a}, who considered since dozen of years magnetization \index{magnetization} reversal by a transfer of spin momentum from the current of spin polarized carriers to localized magnetic moments in ferromagnetic metals. In the case of semiconductors,  the current-induced magnetization \index{magnetization} reversal was demonstrated in submicron pillars of (Ga,Mn)As/GaAs/(Ga,Mn)As \cite{Chiba:2004_b}.  Furthermore, spin-polarized current was shown to displace magnetic domain walls in (Ga,Mn)As with the easy axis perpendicular to the film plane \cite{Yamanouchi:2004_a,Yamanouchi:2006_a}.

\section{Summary and Outlook}
\label{sec:7}

As an outcome of the great progress made in the field of semiconductor
spintronics in the past few years as reviewed above, spin transistors were for the first time
described in The International Technology Roadmap for Semiconductors:
Update 2004 – Emerging Research Devices. Here it was also suggested that spin
transistors might replace unipolar silicon transistors, which have been so
successfully employed since the 1960s. It is, however, also obvious from this review
that a number of challenges are ahead, so that semiconductor spintronics will
attract a lot of attentions of the research community in the years to come.

From the device physics perspective, further works on magnetooptical isolators and modulators
as well as on electrically controlled spin current generation, injection, detection, filtering,
and amplification, particularly in spin bipolar devices \cite{Zutic:2004_a} are expected.
At the same time, further advancement in low-power magnetization \index{magnetization} switching will allow
the development of new generation magnetic random access memories (MRAM) and, perhaps,
extend the use of magnetism towards logics. Last but not least a progress in manipulation of
single electron or nuclear spins in scalable
solid-state devices can be envisaged, though a time scale in question is hard to predict.

Similarly to other branches of condensed matter physics, breakthrough achievements
will be triggered by developments of new materials. Further progress in
p-type doping and magnetic ion incorporation to standard semiconductors
will make it possible to synthesize functional high temperature ferromagnetic
DMS \index{diluted magnetic semiconductors}. At the same time, a control over ferromagnetic precipitates in various
semiconductors
will result in composite materials that will be useful as magnetooptical media and for high density memories.
Particular attention will be paid to insulating ferrimagnetic oxides and
nitrides, which could  serve as spin selective barriers up to well
above room temperature. Moreover, efforts will be undertaken to convert
them into functional magnetic semiconductors by
elaboration of purification methods and mastering doping protocols that will
produce high mobility electrons and holes in these systems. Another line
of research will be devoted to search for nonmagnetic barrier materials in which transmission
coefficients could be electrically adjusted to optimize either reading or writing process
in MRAM cells. Particularly
prospective appear multiferroic systems,
as in these multifunctional materials the coupling between magnetic and electric polarizations offers
new device paradigms.

The important aspect of extensive studies on ferromagnetism in semiconductors discussed in this review is the demonstration of suitability of empirically-constrained theoretical methods in quantitative description of a large body of thermodynamic, micromagnetic, transport, and optical properties of ferromagnetic semiconductors. In particular, a successful description of spintronic effects in both nonmagnetic and magnetic semiconductors is possible provided that all peculiarities of the host band structure, especially those associated with the spin-orbit \index{spin-orbit interaction} interaction, are carefully taken into account. Indeed, as a result of such an effort (Ga,Mn)As has reached the status of the best understood ferromagnet. At the same time, research on DMS \index{diluted magnetic semiconductors} has disclosed shortcomings of today's computational materials science in predicting and elucidating magnetic properties of solids. It appears that this failure of \emph{ab initio} methods (prediction of ferromagnetism in systems where it is absent and inability to explain its nature in materials where it does exist) originates from the co-existence of strong correlation with electronic and magnetic disorder in DMS \index{diluted magnetic semiconductors}. This calls for novel computation protocols that will be able to handle randomness and correlation on equal footing,  also at non-zero temperatures, and will allow for the existence of electronic and/or chemical nanoscale phase separations. Such computational tools, together with advanced methods of spatially resolved material characterization, will in particular answer a persistently raised question on whether a high temperature ferromagnetism is possible in materials containing no magnetic ions.

With no doubt, in course of the years semiconductor spintronics has evolved into an important branch of today's materials science,  condensed matter physics, and device engineering.

\section*{Acknowledgements}
The author would like to thank his numerous Warsaw's co-workers and Hideo Ohno and his co-workers, as indicated in the reference list, for many years of stimulating discussions and fruitful collaboration.

\printindex
\end{document}